\documentclass[12pt,draftclsnofoot,onecolumn]{IEEEtran}
\IEEEoverridecommandlockouts

\usepackage{graphicx}
\usepackage{subfig}
\usepackage{amsmath,amsthm,amsfonts,amssymb}
\usepackage{cite}
\usepackage{bm}
\usepackage{bbm}
\usepackage{url}
\usepackage{array}
\usepackage{color}
\usepackage{multirow}
\usepackage{booktabs}
\usepackage[table,xcdraw]{xcolor}
\usepackage{enumerate}
\usepackage{epstopdf}
\usepackage{tabularx}
\usepackage{algorithm}
\usepackage[english]{babel}
\usepackage{algpseudocode}

\newtheorem{rem}{Remark}
\newtheorem{proposition}{\textbf{Proposition}}

\newenvironment{NewProof}{{\noindent\it Proof.}}{\hfill $\blacksquare$\par}

\algrenewcommand{\algorithmiccomment}[1]{\hfill// #1}

\allowdisplaybreaks[4]

\begin{document}

\title{\makebox[\linewidth]{\parbox{\dimexpr\textwidth+0.5cm\relax}{\centering Reliable Packet Detection for Random Access Networks: Analysis, Benchmark, and Optimization}}}

\author{
Yuyang~Du,~\IEEEmembership{Student Member,~IEEE},
Soung~Chang~Liew,~\IEEEmembership{Fellow,~IEEE},
\thanks{Y. Du and S. C. Liew are with the Department of Information Engineering, The Chinese University of Hong Kong, New Territories, Hong Kong SAR, China (e-mail: \{dy020, soung\}@ie.cuhk.edu.hk). S. C. Liew is the corresponding author.}
}

\maketitle

\vspace{-2em}

\begin{abstract}
Many advanced industrial systems utilize random access in wireless networks to facilitate massive machine communications with burst transmissions. The stringent requirement for ultra-reliability in industrial communication poses a severe challenge for random access: a receiver should neither miss an incoming packet nor get falsely alarmed by noise or interference. Currently, many academic investigations and industry applications rely on the conventional Schmidl-and-Cox (S\&C) algorithm and its variants for packet detection. However, S\&C was originally developed for single-antenna receivers and lacks a rigorous analytical framework for the extension to multi-antenna receiver settings. This paper is a revisit and enhancement of S\&C to fill this gap.  First, we put forth a packet-detection metric called ``compensated autocorrelation", which yields equivalent performance to the S\&C metric but is more analytically tractable. With the new metric, we obtain accurate closed-form expressions for false-alarm and missed-detection probabilities. Second, we introduce the principle of Pareto comparison for packet-detection benchmarking, enabling simultaneous consideration of false alarms and missed detections for a fair comparison between different packet-detection schemes. Third, we experimentally validate that taking the real part of the autocorrelation enhances the performance of S\&C through a new scheme called real-part S\&C (RP-S\&C). Fourth, and perhaps most importantly, the adoption of the new metric, compensated autocorrelation, allows us to extend the single-antenna algorithm to the multi-antenna scenario in a rigorous and analytical manner through a weighted-sum compensated autocorrelation. We formulate two optimization problems, aiming to minimize false-alarm probability and missed-detection probability, respectively. We provide our solutions to these problems along with proofs. We demonstrate through extensive experiments that the optimal weights for false alarms (WFA) is a more desirable scheme than the optimal weights for missed detections (WMD) due to its simplicity, reliability, and superior performance. Our results have significant implications for the design and implementation of packet-detection schemes in random-access networks.
\end{abstract}


\section{Introduction}\label{sec-I}
Random access in wireless networks offers significant benefits for industrial applications such as Industrial Internet-of-Things (IIoT) and sensor networks that thrive on tetherless communication. In contrast to centralized access control, random access enables massive machine communication without a predetermined transmission schedule. For instance, in monitoring applications, a sensor may generate a new packet only upon detecting an anomaly and then transmit this information to a central monitoring station via a wireless channel. As sensor traffic is sporadic, employing random access is more efficient than pre-allocating dedicated wireless resources (e.g., time slots or subcarriers) to each sensor. Moreover, using centralized access control to schedule every sensor becomes impractical when the number of connections surpasses the available wireless resources.

In random access, a receiver does not know when a wireless device will transmit a packet to it. To save power and avoid being wrongly occupied, the packet decoding circuitry of a receiver should not get activated unless a packet is being transmitted, i.e., the receiver needs to detect the incoming packet before decoding it. Therefore, there are two possible causes for reception failures in random access: (i) the packet is not detected; (ii) the packet is detected, but its data cannot be decoded.

Considerable research efforts have been devoted to enhancing the reliability of random access for mission-critical industrial communications \cite{ref1,ref2,ref3}. However, the majority of these studies have primarily focused on packet decoding, while packet detection has received limited attention. Previous packet-detection schemes \cite{ref4,ref5,ref6,ref7,ref8,ref9,ref10} used the conventional Schmidl-and-Cox (S\&C) algorithm \cite{ref11} as the underlying packet-detection scheme. However, a rigorous framework for analyzing the packet detection process is lacking, and the closed-form expressions of missed-detection and false-alarm probabilities in random access are still absent. Further, the existing benchmarking method for packet detection algorithms is defective, as it overlooked the tradeoffs between missed-detection and false-alarm probabilities and focused on minimizing the missed-detection probability as the sole criterion.\footnote{Avoiding getting falsely alarmed is as important as preventing missed detections for three reasons. First, when a false alarm occurs, signal processing circuits are erroneously activated, leading to a decrease in power efficiency. Second, to avoid packet collisions, a random-access device may hold back and refrain from transmitting a packet itself upon encountering a false alarm, resulting in reduced spectrum efficiency. Third, during the false alarm period, as the receiver is occupied decoding the "fake packet", all true incoming packets will not get processed until the receiver realizes the situation and resets its state machine.}

Another seldom-addressed challenge in previous works pertains to the optimization of a packet detection algorithm in the multi-antenna scenario. Conventional S\&C algorithm was originally proposed for single-antenna receivers three decades ago. However, in modern communication systems, receivers are typically equipped with multiple antennas, enabling the possibility of enhancing system reliability via rich spatial diversity (also known as antenna diversity). While previous research efforts have delivered higher decoding reliability by leveraging the spatial diversity \cite{ref12,ref13,ref14}, multi-antenna packet detection has received less attention. Existing studies \cite{ref15,ref16,ref17,ref18,ref19} have extended the S\&C algorithm to the multi-antenna scenario in an \textit{ad hoc} manner due to the lack of a rigorous analytical framework. To the best of our knowledge, no prior research has rigorously analyzed the performance of packet detection in multi-antenna systems, nor has it addressed the optimization challenges associated with such scenarios.

This paper is an attempt to bridge these gaps. We first provide a comprehensive study for the analysis and benchmarking of packet detections in single-antenna random-access systems. After that, we extend our analytical framework to advanced systems with multiple antennas in a rigorous manner and address the optimization problem for multi-antenna packet detection. Our contributions are summarized as follows:

Our first contribution is the proposal of a new metric for packet detection called ``compensated autocorrelation" for single-antenna packet detection, which makes possible a rigorous analytical framework. Previous research mostly used the ratio of autocorrelation and signal power as the packet-detection metric. Rigorous analysis is difficult because the autocorrelation and the signal-power terms contain correlated noises, and their ratio is a complicated function of these correlated noises. The new compensated autocorrelation metric is equivalent to the ratio metric as far as the packet detection performance is concerned. However, the noise characteristic of the compensated autocorrelation is analytically tractable, because the metric contains only a simple summation of correlated noises and can be approximated as a Gaussian random variable. We demonstrate through experiments that our derivations and approximations are precise and reliable. The use of compensated autocorrelation also paves the way for the treatment of packet detection in the multi-antenna scenario (see our fourth contribution below).

Our second contribution is a new benchmarking method. A packet detection algorithm inherently trades off between false alarms and missed detections. Concluding that an algorithm is good simply because of its low missed-detection probability, as is done in many existing papers (e.g., \cite{ref19}), is unreasonable, as that may come at the expense of extremely high false-alarm probability. Our method addresses this problem by introducing Pareto comparison so that we can consider false alarms and missed detections simultaneously.

Our third contribution is the enhancement of the conventional S\&C algorithm. We replace the autocorrelation with its real part and find that our revised scheme, referred to as the real-part S\&C (RP-S\&C), contains less noise than the conventional scheme. We demonstrate the superiority of RP-S\&C over the conventional S\&C.

Our fourth contribution is packet detection in multi-antenna systems building upon the compensated autocorrelation framework. The weighted sum of the individual compensated autocorrelations obtained at different antennas still only contains a sum of noises and is therefore analytically tractable. Using the weighted sum as the metric in the multi-antenna scenario is a natural extension of the single-antenna treatment, and optimality under different criteria can be established rigorously. We consider two specific criteria: (i) minimizing false-alarm probability and (ii) minimizing missed-detection probability. We then give our solutions, the weight assignment for false alarms (WFA) and the weight assignment for missed detections (WMD), to the two optimization problems with rigorous proofs. Last but not least, we discuss implementation details of WFA and WMD and benchmark them under a practical random-access setting with distributed antennas. Based on concrete analyses under practical settings and extensive emulation experiments, we find that WFA is the recommended choice for practical random access due to its simplicity and superior packet-detection performance.

\section{Single-antenna Packet Detection: Analysis, Simulation, and Discussion}\label{sec-II}
\subsection{Conventional S\&C Algorithm and Our Improvement}
\begin{figure}[htbp]
  \centering
  \includegraphics[width=0.45\textwidth]{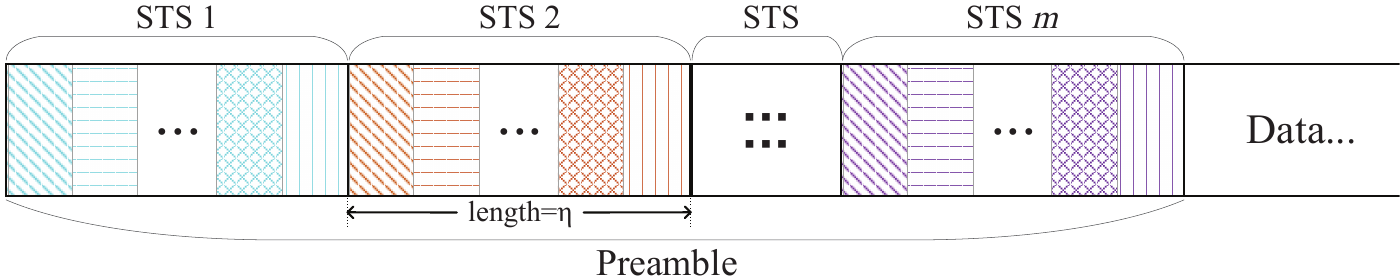}\\
  \captionsetup{font={small}}
  \caption{A general packet format for random access.}
\label{fig:1}
\end{figure}

A random-access system employs repeating sequences to detect packets. Fig. \ref{fig:1} shows a general packet format for random access. The repeating sequences at the beginning of a packet are referred to as short training sequences (STSs), and a collection of multiple STSs forms the preamble sequence. Let us denote the number of STSs by $m$ and the length of each STS by $\eta$. In this paper, for simplicity, we assume that the preamble sequence contains two STSs, i.e., $m=2$. There are several ways to extend the basic treatment here to more general preamble sequences with more than two STSs. That extension will be addressed in a separate paper.

Let the transmitted preamble sequence be $\sqrt {P} s[n]$, where $P$ is the signal power and $s[n]$ is the normalized preamble sequence with index $n$. We have $s[n]=s[n+\eta]$ in the preamble. We can write the average preamble power as
\begin{equation}\label{eq001}
\overline {|s[n]{|^2}} {\rm{ = }}\frac{1}{\eta }\sum\limits_{n = 0}^{\eta  - 1} {|s[n]{|^2}}  = 1
\end{equation}

At the receiver end, the received preamble sequence is
\begin{equation}\label{eq002}
y[n] = \sqrt P s[n] + w[n]
\end{equation}
where $w[n] \sim N(0,{\sigma ^2})$ is the receiver noise. The autocorrelation and average power over the two STSs are
\begin{equation}\label{eq003}
a[n] = \frac{1}{\eta }\sum\limits_{k = 0}^{\eta  - 1} {y[n + k] \cdot } {y^*}[n + \eta  + k]
\end{equation}
\begin{equation}\label{eq004}
b[n] = \frac{1}{{2\eta }}\sum\limits_{k = 0}^{2\eta  - 1} {y[n + k] \cdot } {y^*}[n + k]
\end{equation}

The conventional S\&C algorithm used packet-detection metric $l[n]$ written as
\begin{equation}\label{eq005}
l[n] = \frac{{\left| {a[n]} \right|}}{{b[n]}}
\end{equation}

\begin{figure}[htbp]
  \centering
  \includegraphics[width=0.45\textwidth]{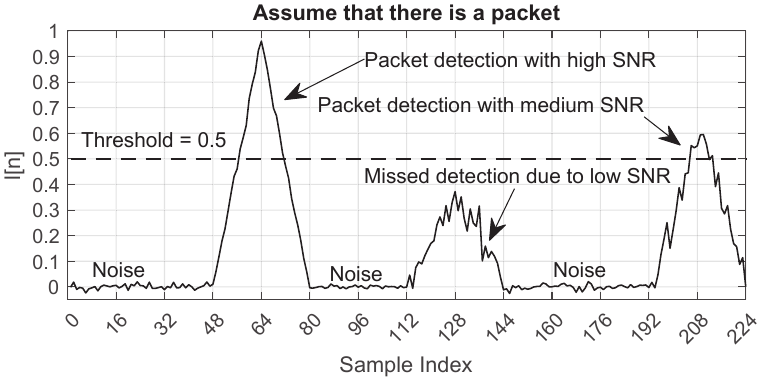}\\
  \captionsetup{font={small}}
  \caption{Illustration of how S\&C detects wireless packets and misses packet detection. Here $\eta=16$.}
\label{fig:2}
\end{figure}

\begin{figure}[htbp]
  \centering
  \includegraphics[width=0.45\textwidth]{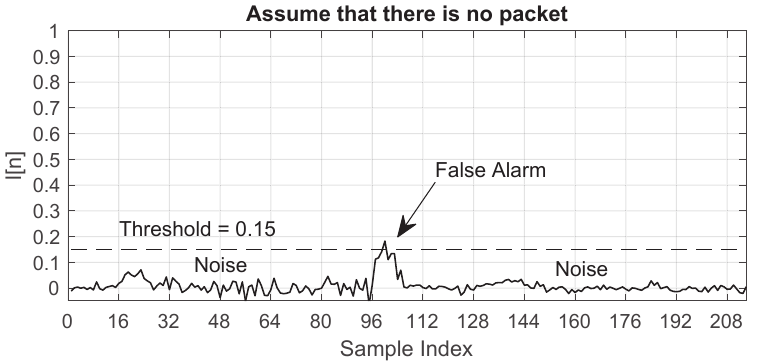}\\
  \captionsetup{font={small}}
  \caption{Illustration of how false alarms occur.}
\label{fig:3}
\end{figure}

Without noise, $l[n]$ reaches its peak value (i.e., $l[n] = 1$) at a particular index $n$ corresponding to the beginning of the first preamble sample. With noise, on the other hand, $l[n]$ is in general smaller than one. S\&C compares $l[n]$ with a pre-defined threshold $\rho $. Fig. \ref{fig:2} and Fig. \ref{fig:3} illustrate the packet-detection process.

In Fig. \ref{fig:2}, we assume that there are three packets. For an incoming packet, if the peak of its $l[n]$ is larger than threshold $\rho $ (e.g., the first and the last $l[n]$ peak in Fig. \ref{fig:2}), then the receiver declares a packet is detected and this triggers the subsequent signal processing to decode the packet. Otherwise, if the peak value is smaller than $\rho $ (e.g., the second $l[n]$ peak in Fig. \ref{fig:2}), the receiver performs no action, and an event of missed packet detection occurs. Packet missed detections may occur often when the antenna signal-to-noise ratio (SNR) is too low or the threshold $\rho $ is set to too high.

In Fig. \ref{fig:3}, we assume there is no packet (i.e., pure noise input). We see that $l[n]$ is very close to zero because the input noise is random. Nevertheless, we may still observe a $l[n]$ larger than $\rho$, and when that occurs, a false alarm event occurs. In general, false alarms are more likely to occur the lower the threshold $\rho$. Hence, adjusting the value of $\rho$ amounts to trading off missed-detection performance and false-alarm performance.

In (\ref{eq005}), the absolute value of $a[n]$ is taken because a large carrier frequency offset (CFO) can disperse a relatively huge amount of signal power into the imaginary part of $a[n]$. In the absence of noise and CFO, on the other hand, $a[n]$ is real. Thanks to advancements in hardware and semiconductor technology in the past 30 years, modern communication systems have a much lower CFO than old systems developed at the time when S\&C algorithm was proposed \cite{ref20,ref21}. With negligible CFO, in the presence of noise, the signal is entirely contained in the real part of $a[n]$, and the imaginary part of $a[n]$ consists of noise only. By taking the absolute value of $a[n]$, S\&C algorithm inadvertently includes much noise in $l[n]$. As we will justify in Section \ref{sec-III}, under the weak CFO condition of a modern communication system, taking the real part of $a[n]$ can enhance the packet-detection performance. Unless stated otherwise, the rest of this paper uses an alternative metric $l_R[n]$ written as
\begin{equation}\label{eq006}
{l_R}[n] = \frac{{{a_R}[n]}}{{b[n]}}
\end{equation}
where the subscript $R$ represents the real part of a variable. We refer to the modification we made over the conventional S\&C algorithm as the real-part S\&C (RP-S\&C) algorithm.

For packet detection, we are interested in whether $l_R[n]>\rho$.
Yet, analyzing $l_R[n]$ (also, the original $l[n]$) is challenging as it is the ratio of two non-independent random variables. We note that saying $l_R[n]>\rho$ is equivalent to saying
\begin{equation}\label{eq007}
r[n] \buildrel \Delta \over = {a_R}[n] - \rho b[n]{\kern 1pt} {\kern 1pt} {\kern 1pt} {\kern 1pt}  > 0
\end{equation}

We refer to $r[n]$ as the ``compensated autocorrelation". That is, $a_R[n]$ is the real part of the autocorrelation and we compensate it by subtracting $\rho b[n]$ from it and checking whether the resulting value is larger than 0. Having a large $a_R[n]$ does not necessarily mean that there is a packet because it could be due to a large exogenous interference (e.g., Bluetooth interference on WiFi). However, large exogenous interference also has relatively larger $b[n]$ compared with $a_R[n]$, and hence $r[n]$ is likely to be small in that case.

There are two key advantages of focusing on $r[n]$ rather than $l_R[n]$. First, compared with $l_R[n]$, $r[n]$ is much easier to analyze. We know that $a_R[n]$ and $b[n]$ contain correlated noises (see (\ref{eq009}) and (\ref{eq015}) below), and the noise in $l_R[n]$ is a complicated function of these correlated noises. The noise in $r[n]$, on the other hand, consists of the simple summation of these correlated noises because it is a simple linear combination of $a_R[n]$ and $b[n]$. For a practical preamble, we can approximate $r[n]$ as a Gaussian random variable in our analysis (we will elaborate later). We show in Subsection B that it is easy to compute the mean and variance of $r[n]$ and approximate $r[n]$ as a Gaussian random variable with that mean and variance.

Second, for a multi-antenna system, we could add the weighted $r[n]$ of different antennas to form a weighted-combined $r[n]$ and compare that with a threshold for packet detection purposes. Again, the weighted-combined $r[n]$ is amenable to analysis since the weighted combination can also be approximated as a Gaussian random variable. This allows us to investigate the optimality of different weight combinations on a rigorous basis.

\subsection{Analysis of RP-S\&C with Gaussian Approximation}
In this subsection, we first analyze $b[n]$, ${a_R}[n]$, and their cross term ${a_R}[n] b[n]$. Based on these analyses, we obtain the mean and variance of $r[n]$. We then approximate $r[n]$ as a Gaussian random variable with the computed mean and variance. Finally, we utilize the distribution of $r[n]$ to determine the false-alarm probability and the missed-detection probability.

Let us start with $b[n]$. We have that
\begin{equation}\label{eq008}
E\left( {b[n]} \right) = P + {\sigma ^2}
\end{equation}

For the analysis of $Var\left( {b[n]} \right)$, we write $b[n]$ as
\begin{equation}\label{eq009}
\small
\begin{aligned}
b[n]
&= \frac{1}{{2\eta }}\sum\limits_{k = 0}^{2\eta  - 1} {y[n + k] \cdot {y^*}[n + k]} \\
&= \frac{1}{{2\eta }}\sum\limits_{k = 0}^{2\eta  - 1} {\left( {P|s[n + k]{|^2} + \sqrt P s[n + k]{w^*}[n + k] + \sqrt P {s^*}[n + k]w[n + k] + w[n + k]{w^*}[n + k]} \right)} \\
&= P + \frac{{\sqrt P }}{\eta }\sum\limits_{k = 0}^{2\eta  - 1} {\left( {{s_R}[n + k]{w_R}[n + k] + {s_I}[n + k]{w_I}[n + k]} \right)}  + \frac{1}{{2\eta }}\sum\limits_{k = 0}^{2\eta  - 1} {\left( {w_R^2[n + k] + w_I^2[n + k]} \right)}
\end{aligned}
\normalsize
\end{equation}
where subscript $R$ and subscript $I$ represent the real and imaginary parts of a term, respectively.

Now, we have (note that $E\left( {w_R^3[n]} \right) = E\left( {w_I^3[n]} \right) = 0$  in the following derivation)
\begin{equation}\label{eq010}
\small
\begin{aligned}
E
&\left( {{b^2}[n]} \right)\\
&= E\left\{ {{{\left( {P + \frac{{\sqrt P }}{\eta }\sum\limits_{k = 0}^{2\eta  - 1} {\left( {{s_R}[n + k]{w_R}[n + k] + {s_I}[n + k]{w_I}[n + k]} \right)}  + \frac{1}{{2\eta }}\sum\limits_{k = 0}^{2\eta  - 1} {\left( {w_R^2[n + k] + w_I^2[n + k]} \right)} } \right)}^2}} \right\}\\
&= \left( {{P^2} + 2 P {\sigma ^2} + \frac{P}{\eta }  {\sigma ^2}} \right) + \frac{{E\left( {w_R^4[n]} \right) + E\left( {w_I^4[n]} \right) + 2E\left( {w_R^2[n]} \right) E\left( {w_I^2[n]} \right)}}{{2\eta }} +  \frac{{4(2\eta  - 1)  E\left( {w_R^2[n]} \right) \cdot E\left( {w_I^2[n]} \right)}}{{2\eta }}\\
&= \left( {{P^2} + \frac{{2\eta  + 1}}{\eta }P{\sigma ^2}} \right) + \frac{{E\left( {w_R^4[n]} \right)}}{\eta } + \frac{{{E^2}\left( {w_R^2[n]} \right)}}{\eta } + \frac{{2(2\eta  - 1){E^2}\left( {w_R^2[n]} \right)}}{\eta }\\
&= \left( {{P^2} + \frac{{2\eta  + 1}}{\eta }P{\sigma ^2}} \right) + \frac{{E\left( {w_R^4[n]} \right)}}{\eta } + \frac{{{{{\sigma ^4}} \mathord{\left/
 {\vphantom {{{\sigma ^4}} 4}} \right.
 \kern-\nulldelimiterspace} 4} + 2({{2\eta  - 1){\sigma ^4}} \mathord{\left/ {\vphantom {{2\eta  - 1){\sigma ^4}} 4}} \right. \kern-\nulldelimiterspace} 4}}}{\eta }
\end{aligned}
\normalsize
\end{equation}

The random variable $\sqrt {{2 \mathord{\left/ {\vphantom {2 {{\sigma ^2}}}} \right.\kern-\nulldelimiterspace} {{\sigma ^2}}}} {w_R}$ is of the standard normal distribution. Thus, $({2 \mathord{\left/ {\vphantom {2 {{\sigma ^2}}}} \right. \kern-\nulldelimiterspace} {{\sigma ^2}}})w_R^2[n]$ is a chi-square distribution of degree 1, and we have
\begin{equation}\label{eq011}
E\left[ {{{\left( {\frac{2}{{{\sigma ^2}}}w_R^2[n]} \right)}^2}} \right]
= Var\left( {\frac{2}{{{\sigma ^2}}}w_R^2[n]} \right) + {E^2}\left( {\frac{2}{{{\sigma ^2}}}w_R^2[n]} \right) = 3\\
{\kern 1pt} {\kern 1pt} {\kern 1pt}{\kern 1pt} {\kern 1pt} {\kern 1pt} \Rightarrow {\kern 1pt} {\kern 1pt} {\kern 1pt} {\kern 1pt}  E\left( {w_R^4[n]} \right) = 3{\sigma ^4}/4
\end{equation}

Substituting (\ref{eq011}) into (\ref{eq010}), we have
\begin{equation}\label{eq012}
E\left( {{b^2}[n]} \right)=\left( {{P^2}{\rm{ + 2}}P{\sigma ^2} + {\sigma ^4}} \right) + \frac{{2P{\sigma ^2} + {\sigma ^4}}}{{2\eta }}
\end{equation}

With (\ref{eq012}), we have
\begin{equation}\label{eq013}
Var\left( {b[n]} \right) = E\left( {{b^2}[n]} \right){\kern 1pt} {\kern 1pt} {\kern 1pt}  - {E^2}\left( {b[n]} \right){\kern 1pt} {\kern 1pt} {\kern 1pt}  = \frac{{2P{\sigma ^2} + {\sigma ^4}}}{{2\eta }}
\end{equation}

We next look at ${a_R}[n]$. We can write $a[n]$ as
\begin{equation}\label{eq014}
\small
\begin{aligned}
a[n]
&= \frac{1}{\eta }\sum\limits_{k = 0}^{\eta  - 1} {y[n + k] \cdot } {y^*}[n + \eta  + k]\\
&= \frac{1}{\eta }\sum\limits_{k = 0}^{\eta  - 1} {\left( {\sqrt P s[n + k] + w[n + k]} \right) \cdot \left( {\sqrt P {s^{\rm{*}}}[n + k] + {w^{\rm{*}}}[n + \eta  + k]} \right)} \\
&= P + \frac{1}{\eta }\sum\limits_{k = 0}^{\eta {\rm{ - }}1} {\left\{ {\sqrt P \left( {s[n + k]{w^*}[n + \eta  + k] + {s^*}[n + k]w[n + k]} \right) + w[n + k]{w^*}[n + \eta  + k]} \right\}}
\end{aligned}
\normalsize
\end{equation}

We extract the real part of (\ref{eq014}) and write $a_R[n]$ as
\begin{equation}\label{eq015}
{a_R}[n] = P + {\kern 1pt} \frac{1}{\eta }\sum\limits_{k = 0}^{\eta  - 1} {\left\{ \begin{array}{l}
\sqrt P \left[ \begin{array}{l}
\left( {{s_R}[n + k]{w_R}[n + \eta  + k] + {s_I}[n + k]{w_I}[n + \eta  + k]} \right)\\
{\kern 1pt} {\kern 1pt} {\kern 1pt} {\kern 1pt} {\kern 1pt} {\kern 1pt} {\kern 1pt} {\kern 1pt} {\kern 1pt} {\kern 1pt} {\kern 1pt} {\kern 1pt} {\kern 1pt} {\kern 1pt} {\kern 1pt} {\kern 1pt} {\kern 1pt}  + \left( {{s_R}[n + k]{w_R}[n + k] + {s_I}[n + k]{w_I}[n + k]} \right)
\end{array} \right]\\
{\kern 1pt} {\kern 1pt} {\kern 1pt} {\kern 1pt}{\kern 1pt} {\kern 1pt} {\kern 1pt} {\kern 1pt} {\kern 1pt} {\kern 1pt} {\kern 1pt} {\kern 1pt}  + {w_R}[n + k]{w_R}[n + \eta  + k] + {w_I}[n + k]{w_I}[n + \eta  + k]
\end{array} \right\}}
\end{equation}

From (\ref{eq015}), by exploiting the fact that the zero-mean Gaussian noise terms at different time indexes are independent, we have that
\begin{equation}\label{eq016}
E\left( {{a_R}[n]} \right) = P
\end{equation}

Similarly, we have
\begin{equation}\label{eq017}
\small
\begin{aligned}
E\left( {a_R^2[n]} \right)
&= {P^2} + \frac{1}{{{\eta ^2}}}\sum\limits_{k = 0}^{\eta  - 1} {\left\{ \begin{array}{l}
P  E\left( \begin{array}{l}
\left( {s_R^2[n + k]w_R^2[n + \eta  + k] + s_I^2[n + k]w_I^2[n + \eta  + k]} \right)\\
{\kern 1pt} {\kern 1pt} {\kern 1pt} {\kern 1pt} {\kern 1pt} {\kern 1pt} {\kern 1pt} {\kern 1pt} {\kern 1pt} {\kern 1pt} {\kern 1pt} {\kern 1pt} {\kern 1pt} {\kern 1pt} {\kern 1pt} {\kern 1pt} {\kern 1pt}  + \left( {s_R^2[n + k]w_R^2[n + k] + s_I^2[n + k]w_I^2[n + k]} \right)
\end{array} \right)\\
{\kern 1pt} {\kern 1pt} {\kern 1pt} {\kern 1pt} {\kern 1pt} {\kern 1pt} {\kern 1pt} {\kern 1pt} {\kern 1pt} {\kern 1pt} {\kern 1pt}  + E\left( {w_R^2[n + k]w_R^2[n + \eta  + k] + w_I^2[n + k]w_I^2[n + \eta  + k]} \right)
\end{array} \right\}}\\
&= {P^2} + \frac{{P{\sigma ^2}}}{\eta } + \frac{{{\sigma ^4}}}{{2\eta }}
\end{aligned}
\normalsize
\end{equation}

From (\ref{eq016}) and (\ref{eq017}), we get
\begin{equation}\label{eq018}
Var({a_R}[n]) = E(a_R^2[n]) - {E^2}({a_R}[n]) = \frac{{2P{\sigma ^2} + {\sigma ^4}}}{{2\eta }}
\end{equation}

We next look at ${a_R}[n] b[n]$. With (\ref{eq009}) and (\ref{eq015}), we have
\begin{equation}\label{eq019}
\small
\begin{array}{l}
E\left( {{a_R}[n] b[n]} \right){\kern 1pt} \\
{\kern 1pt} {\kern 1pt} {\kern 1pt} {\kern 1pt}  = E\left\{ \begin{array}{l}
\left( {P + {\kern 1pt} \frac{{\sqrt P }}{\eta }\sum\limits_{k = 0}^{\eta  - 1} {\left[ \begin{array}{l}
\left( \begin{array}{l}
{s_R}[n + k]{w_R}[n + \eta  + k]\\
 + {s_I}[n + k]{w_I}[n + \eta  + k]
\end{array} \right)\\
{\kern 1pt} {\kern 1pt}  + \left( \begin{array}{l}
{s_R}[n + k]{w_R}[n + k]\\
 + {s_I}[n + k]{w_I}[n + k]
\end{array} \right)
\end{array} \right]}  + \frac{1}{\eta }\sum\limits_{k = 0}^{\eta  - 1} {\left[ \begin{array}{l}
{w_R}[n + k]\\
{\kern 1pt} {\kern 1pt} {\kern 1pt}  \cdot {w_R}[n + \eta  + k]\\
 + {w_I}[n + k]\\
{\kern 1pt} {\kern 1pt} {\kern 1pt}  \cdot {w_I}[n + \eta  + k]
\end{array} \right]} } \right){\kern 1pt} {\kern 1pt} {\kern 1pt} {\kern 1pt} \\
{\kern 1pt} {\kern 1pt} {\kern 1pt} {\kern 1pt} {\kern 1pt} {\kern 1pt} {\kern 1pt} {\kern 1pt} {\kern 1pt} {\kern 1pt} {\kern 1pt} {\kern 1pt} {\kern 1pt} {\kern 1pt} {\kern 1pt} {\kern 1pt} {\kern 1pt}  \times \left( \begin{array}{l}
P + \frac{{\sqrt P }}{\eta }\sum\limits_{k = 0}^{2\eta  - 1} {\left( {{s_R}[n + k]{w_R}[n + k] + {s_I}[n + k]{w_I}[n + k]} \right)} \\
{\kern 1pt} {\kern 1pt} {\kern 1pt} {\kern 1pt} {\kern 1pt} {\kern 1pt} {\kern 1pt} {\kern 1pt} {\kern 1pt} {\kern 1pt}  + \frac{1}{{2\eta }}\sum\limits_{k = 0}^{2\eta  - 1} {\left( {w_R^2[n + k] + w_I^2[n + k]} \right)}
\end{array} \right)
\end{array} \right\}\\
{\kern 1pt} {\kern 1pt} {\kern 1pt} {\kern 1pt}  = {P^2} + \frac{P}{{{\eta ^2}}}\sum\limits_{k = 0}^{\eta  - 1} {\left\{ \begin{array}{l}
s_R^2[n + k]E(w_R^2[n + k])\\
 + s_I^2[n + k]E(w_R^2[n + k])
\end{array} \right\}} {\kern 1pt} \\
{\kern 1pt} {\kern 1pt} {\kern 1pt} {\kern 1pt} {\kern 1pt} {\kern 1pt} {\kern 1pt} {\kern 1pt} {\kern 1pt} {\kern 1pt} {\kern 1pt} {\kern 1pt} {\kern 1pt} {\kern 1pt} {\kern 1pt} {\kern 1pt} {\kern 1pt} {\kern 1pt} {\kern 1pt} {\kern 1pt} {\kern 1pt} {\kern 1pt} {\kern 1pt} {\kern 1pt} {\kern 1pt} {\kern 1pt} {\kern 1pt} {\kern 1pt} {\kern 1pt}  + \frac{P}{{{\eta ^2}}}\sum\limits_{k = 0}^{\eta  - 1} {\left\{ \begin{array}{l}
s_R^2[n + k]E(w_R^2[n + \eta  + k])\\
 + s_I^2[n + k]E(w_R^2[n + \eta  + k])
\end{array} \right\}}  + \frac{P}{{2\eta }}\sum\limits_{k = 0}^{2\eta  - 1} {\left\{ \begin{array}{l}
E(w_R^2[n + k])\\
 + E(w_I^2[n + k])
\end{array} \right\}} \\
{\kern 1pt} {\kern 1pt} {\kern 1pt} {\kern 1pt}  = {P^2} + \frac{{P{\sigma ^2}}}{\eta } + P{\sigma ^2}
\end{array}
\normalsize
\end{equation}

With the above analyses of $b[n]$ and ${a_R}[n]$, we calculate $E\left( {r[n]} \right)$ as
\begin{equation}\label{eq020}
E\left( {r[n]} \right) = E\left( {{a_R}[n]{\kern 1pt} } \right) - \rho E\left( {b[n]{\kern 1pt} } \right) = (1 - \rho )P - \rho {\sigma ^2}
\end{equation}

Further, with (\ref{eq012}), (\ref{eq017}), and (\ref{eq019}), we have
\begin{equation}\label{eq021}
\begin{aligned}
E\left( {{r^2}[n]} \right)
&= E\left( {a_R^2[n]} \right){\rm{ + }}{\rho ^2}E\left( {{b^2}[n]} \right) - 2\rho E\left( {{a_R}[n]b[n]{\kern 1pt} } \right)\\
&= {\left( {1 - \rho } \right)^2}{P^2} + \frac{{{{\left( {1 - \rho } \right)}^2} + 2\eta \rho \left( {\rho  - 1} \right)}}{\eta }P + \frac{{1 + {\rho ^2}\left( {2\eta  + 1} \right)}}{{2\eta }}{\sigma ^4}
\end{aligned}
\end{equation}

With (\ref{eq020}), and (\ref{eq021}), we have
\begin{equation}\label{eq022}
Var(r[n]) = E\left( {{r^2}[n]} \right) - {E^2}\left( {r[n]} \right) = \frac{{{{(1 - \rho )}^2}}}{\eta }P{\sigma ^2} + \frac{{1 + {\rho ^2}}}{{2\eta }}{\sigma ^4}
\end{equation}

We now explain why $r[n]$ can be approximated as a Gaussian random variable. From (\ref{eq009}), we see that $b[n]$ is an average of multiple terms. In a practical random-access system, the number of terms in $b[n]$ can be quite large (see Subsection C for justifications). Hence, we can apply the Central Limit Theorem to approximate $b[n]$ as a Gaussian variable \cite{CentralLimitTheorem}. Similarly, with the expression of $a_R[n]$ in ((\ref{eq015}), we can approximate $a_R[n]$ as Gaussian by the same reasoning. Thus, overall, $r[n]$ can be Gaussian approximated, since it is a linear combination of $a_R[n]$ and $b[n]$. With (\ref{eq021}), (\ref{eq022}), and the Gaussian approximations, we can write the distribution of $r[n]$ as
\begin{equation}\label{eq023}
r[n] \sim N\left( {(1 - \rho )P - \rho {\sigma ^2},{\kern 1pt} {\kern 1pt} {\kern 1pt} {\kern 1pt} \frac{{{{(1 - \rho )}^2}}}{\eta }P{\sigma ^2} + \frac{{1 + {\rho ^2}}}{{2\eta }}{\sigma ^4}} \right)
\end{equation}

We can define things in terms of SNR by transforming $r[n]$ to ${{r[n]} \mathord{\left/ {\vphantom {{r[n]} {{\sigma ^2}}}} \right. \kern-\nulldelimiterspace} {{\sigma ^2}}}$. After the transformation, we have
\begin{equation}\label{eq024}
r[n] \sim N\left( {(1 - \rho )\gamma  - \rho {\kern 1pt} {\kern 1pt} {\kern 1pt} {\kern 1pt} ,{\kern 1pt} {\kern 1pt} {\kern 1pt} {\kern 1pt} {\kern 1pt} {\kern 1pt} {\kern 1pt} \frac{{{{(1 - \rho )}^2}}}{\eta }\gamma  + \frac{{1 + {\rho ^2}}}{{2\eta }}} \right)
\end{equation}
where $\gamma = {P \mathord{\left/ {\vphantom {P {{\sigma ^2}}}} \right. \kern-\nulldelimiterspace} {{\sigma ^2}}}$ is the SNR of the studied antenna. In the rest of this paper, unless stated otherwise, we mean the post-transformation $r[n]$ written as the function of $\gamma$ when we mention $r[n]$.

Note that $r[n]$ in (\ref{eq024}) represents the general setting. To analyze the missed-detection probability, we need to assume that there is a packet over the air (i.e., $\gamma  \ne 0$). To analyze the false-alarm probability, on the other hand, we need to assume that there is no packet and there is only noise (i.e., $\gamma  = 0$). For clear descriptions, let us distinguish $r[n]$ with ``packet and noise input" and ``noise input only" with subscript $P$ and subscript $N$ (i.e., ${r_P}[n]$ and ${r_N}[n]$), respectively. Furthermore, we use $r[n]$ to represent general cases regardless of the input type.

Now, we have
\begin{equation}\label{eq025}
{r_P}[n] \sim N\left( {(1 - \rho )\gamma  - \rho {\kern 1pt} {\kern 1pt} {\kern 1pt} {\kern 1pt} ,{\kern 1pt} {\kern 1pt} {\kern 1pt} {\kern 1pt} {\kern 1pt} {\kern 1pt} {\kern 1pt} \frac{{{{(1 - \rho )}^2}}}{\eta }\gamma  + \frac{{1 + {\rho ^2}}}{{2\eta }}} \right),{\kern 1pt} {\kern 1pt} {\kern 1pt} {\kern 1pt} {\kern 1pt} {\kern 1pt} {\kern 1pt} {\kern 1pt} \gamma  \ne 0
\end{equation}

Substituting $\gamma=0$ into (\ref{eq024}), we have
\begin{equation}\label{eq026}
{r_N}[n]\sim N\left( { - \rho {\kern 1pt} {\kern 1pt} {\kern 1pt} {\kern 1pt} ,{\kern 1pt} {\kern 1pt} {\kern 1pt} {\kern 1pt} {\kern 1pt} {\kern 1pt} {\kern 1pt} \frac{{1 + {\rho ^2}}}{{2\eta }}} \right)
\end{equation}

To analyze false alarms and missed detections, we define $z$ to be the normalized $r[n]$:
\begin{equation}\label{eq027}
z = \frac{{r[n] - E\left( {r[n]} \right)}}{{\sqrt {Var\left( {r[n]} \right)} }},{\kern 1pt} {\kern 1pt} {\kern 1pt} {\kern 1pt} {\kern 1pt} {\kern 1pt} {\kern 1pt} {\kern 1pt} {\kern 1pt} {\kern 1pt} {\kern 1pt} {\kern 1pt} {\kern 1pt} {\kern 1pt} {\kern 1pt} {\kern 1pt} {\kern 1pt} {\kern 1pt} {\kern 1pt} {\kern 1pt} {\kern 1pt} {\kern 1pt} {\kern 1pt} {\kern 1pt} z \sim N(0,1)
\end{equation}

Note the normalization in (\ref{eq027}) is the general case that works for both $r_P[n]$ and $r_N[n]$.

We now assume that there is a packet. By the definition of missed detection (i.e., not claiming packet detection when there is a packet), we write the missed-detection probability $P_{MD}$ as
\begin{equation}\label{eq028}
\begin{array}{l}
{P_{MD}} = \frac{1}{{\sqrt {2\pi } }}\int_{ - \infty }^{ - \frac{{E\left( {{r_P}[n]} \right)}}{{\sqrt {Var\left( {{r_P}[n]} \right)} }}} {{\kern 1pt} {\kern 1pt} {\kern 1pt} {e^{ - \frac{{{z^2}}}{2}}}} dz = \frac{1}{{\sqrt {2\pi } }}\int_{\frac{{E\left( {{r_P}[n]} \right)}}{{\sqrt {Var\left( {{r_P}[n]} \right)} }}}^\infty  {{\kern 1pt} {\kern 1pt} {\kern 1pt} {e^{ - \frac{{{z^2}}}{2}}}} dz\\
{\kern 1pt}{\kern 1pt}{\kern 1pt}{\kern 1pt} {\kern 1pt} {\kern 1pt} {\kern 1pt} {\kern 1pt} {\kern 1pt} {\kern 1pt} {\kern 1pt} {\kern 1pt} {\kern 1pt} {\kern 1pt} {\kern 1pt} {\kern 1pt} {\kern 1pt} {\kern 1pt} {\kern 1pt} {\kern 1pt} {\kern 1pt} {\kern 1pt} {\kern 1pt} {\kern 1pt}  = Q\left( {\frac{{E\left( {{r_P}[n]} \right)}}{{\sqrt {Var\left( {{r_P}[n]} \right)} }}} \right) = Q\left( {\frac{{\sqrt \eta  \left( {(1 - \rho )\gamma  - \rho } \right){\kern 1pt} }}{{\sqrt {{\kern 1pt} {{(1 - \rho )}^2}\gamma  + {{(1 + {\rho ^2})} \mathord{\left/
 {\vphantom {{(1 + {\rho ^2})} 2}} \right.
 \kern-\nulldelimiterspace} 2}} }}} \right)
\end{array}
\end{equation}
where $Q(.)$ is the well-known Q function \cite{BetaHandBook}.

We derive the false-alarm probability by assuming that there is no packet. By the definition of false alarm (i.e., claiming packet detection when there is no packet), we write the false-alarm probability $P_{FA}$ as
\begin{equation}\label{eq029}
{P_{FA}} = \frac{1}{{\sqrt {2\pi } }}\int_{ - \frac{{E\left( {{r_N}[n]} \right)}}{{\sqrt {Var\left( {{r_N}[n]} \right)} }}}^\infty  {{\kern 1pt} {\kern 1pt} {\kern 1pt} {\kern 1pt} {e^{ - \frac{{{x^2}}}{2}}}} dx = Q\left( { - \frac{{E\left( {{r_N}[n]} \right)}}{{\sqrt {Var\left( {{r_N}[n]} \right)} }}} \right) = Q\left( {\sqrt {\frac{{2\eta {\rho ^2}}}{{1 + {\rho ^2}}}} } \right)
\end{equation}

\subsection{Simulations and Discussions}
\begin{figure*}[htbp]
  \centering
  \includegraphics[width=0.9\textwidth]{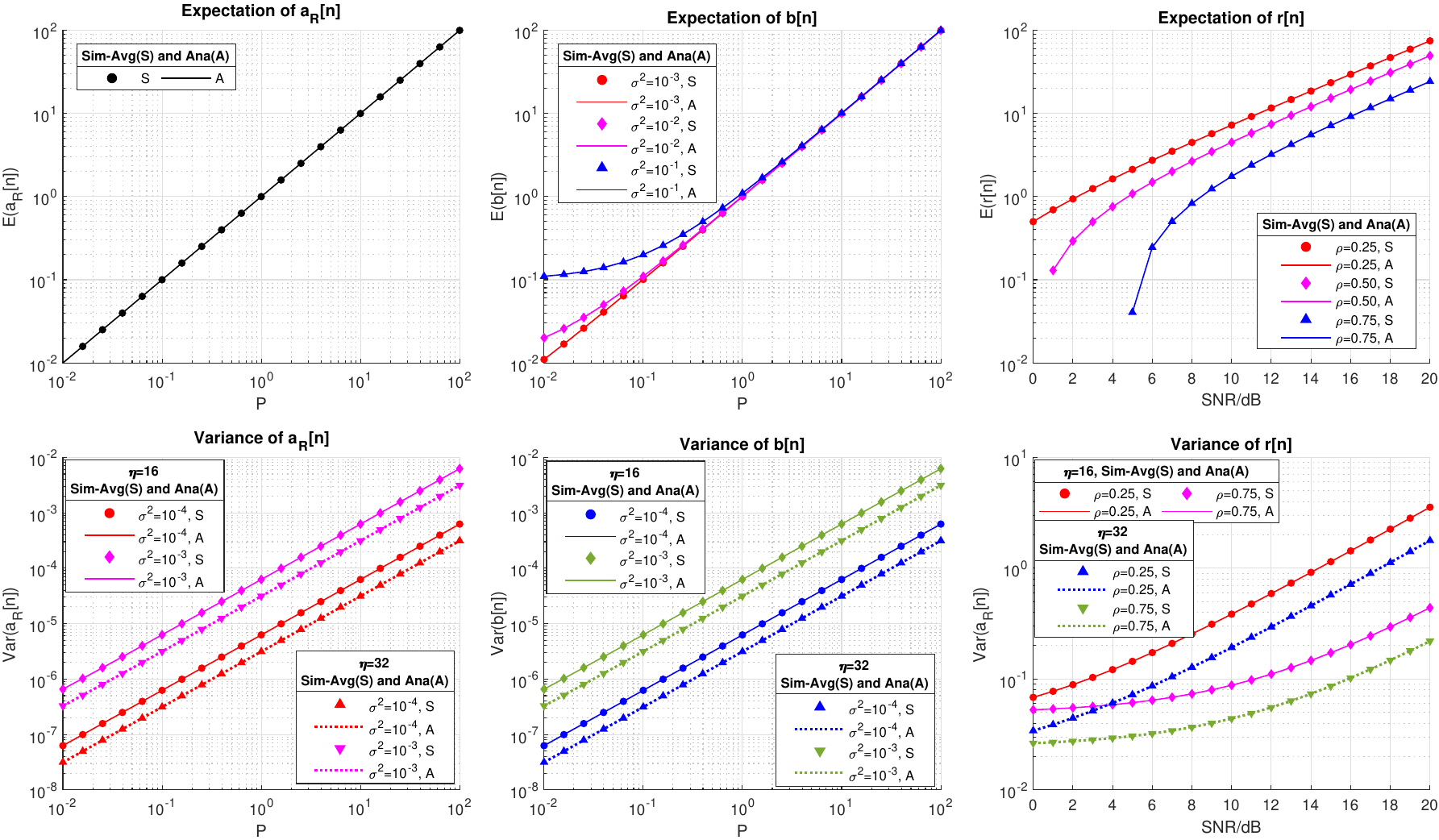}\\
  \captionsetup{font={small}}
  \caption{Simulated and analytical results for the expectations and variances of $a_R[n]$, $b[n]$, and $r[n]$. Here $P$ is the signal power, and $\sigma^2$ is the noise power (see (\ref{eq002}) for the general expression of the received signal). Analytical curves for $a_R[n]$ are plotted according to (\ref{eq016}) and (\ref{eq018}); curves for $b[n]$ are plotted according to (\ref{eq008}) and (\ref{eq013}). The $r[n]$ here is the post-transformation $r[n]$ in (\ref{eq024})}.
\label{fig:4}
\end{figure*}

This subsection validates our derivations and the Gaussian assumptions in subsection B through simulations. Fig. \ref{fig:4} compares the simulated results of $a_R[n]$, $b[n]$, and $r[n]$ with our analytical results under various noise and threshold settings. We conducted multiple simulations and averaged the results to eliminate the randomness in individual simulations. The analytical curves are plotted based on the expressions in subsection B. As the figure shows, the simulated results closely align with the analytical expressions, affirming the correction of our derivations.

\begin{figure}[htbp]
  \centering
  \subfloat[]  {\label{fig:5} \includegraphics[width=0.45\textwidth]{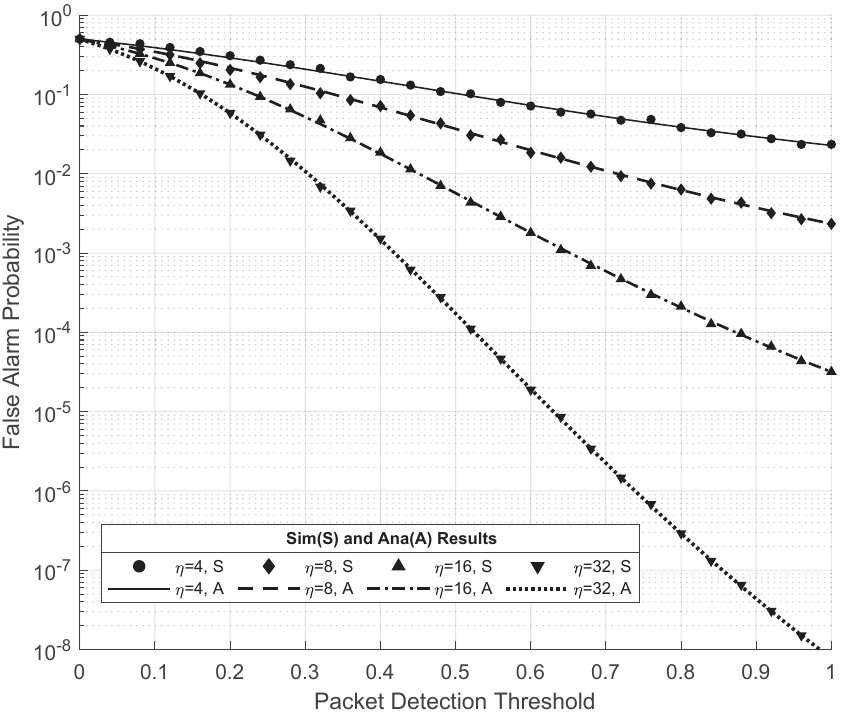}}
  \quad \quad
  \subfloat[]  {\label{fig:6} \includegraphics[width=0.45\textwidth]{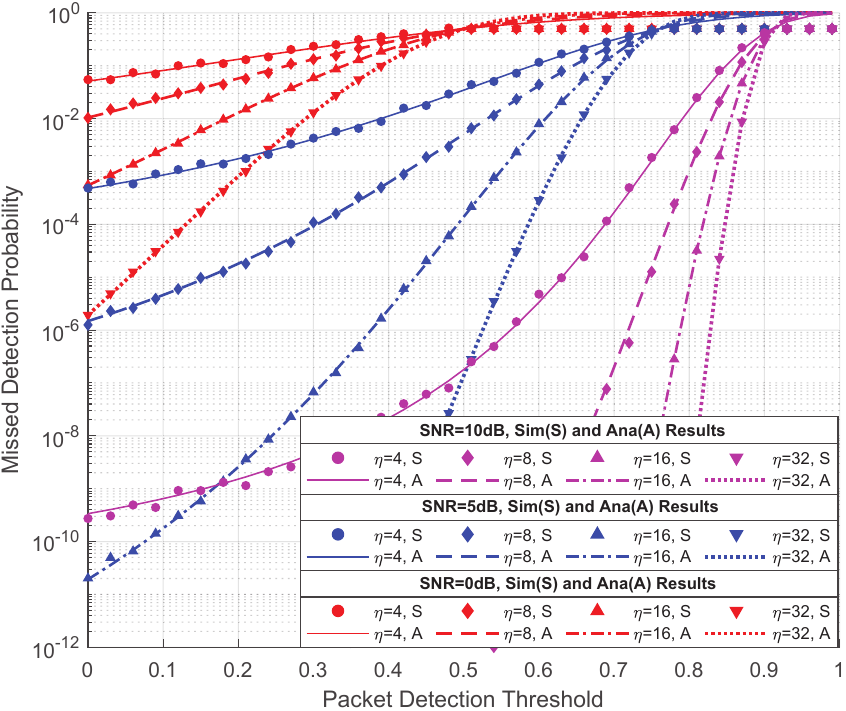}}
  \captionsetup{font={small}}
  \caption{(a) Simulated and analytical results of false-alarm probability under various $\eta$ and $\rho$ settings. (b) Simulated and analytical results of missed-detection probability under various $\eta$, $\gamma$, and $\rho$ settings.}
\end{figure}

We next simulate ${P_{MD}}$ and ${P_{FA}}$. Recall that we made Gaussian approximations on $a_R[n]$, $b[n]$, and their linear combination $r[n]$ to write ${P_{MD}}$ and ${P_{FA}}$ in the form of the Q function. We can see from Fig. \ref{fig:5} and Fig. \ref{fig:6} that the simulated results well match our analysis when there are no less than 16 terms in the summation of $a_R[n]$ and $b[n]$. i.e., $\eta  \ge 16$. This is because the approximation of Central Limit Theorem can be very precise when the number of terms in the summation is large. In a practical random-access system, the length of an STS is typically no smaller than 16. For example, in IEEE 802.11, the total preamble length is 160 samples \cite{IEEE80211}. If we apply the two-STS setting as in this paper, we have $\eta  = 80$. Hence, we can confidently say that our Gaussian approximations on $a_R[n]$, $b[n]$ and $r[n]$ can be very precise on realistic random-access systems, and the expressions of ${P_{MD}}$ and ${P_{FA}}$ we derived in (\ref{eq028}) and (\ref{eq029}) are trustworthy.

\section{Benchmarking RP-S\&C with Conventional S\&C}\label{sec-III}
This section proposes a rigorous method to benchmark the performance of various packet detection schemes. We use this method to compare the performance of the conventional S\&C algorithm with our RP-S\&C algorithm.

Previous studies have evaluated packet detection schemes solely based on the missed-detection probability. For instance, \cite{ref19} experimentally investigated several packet detection schemes for vehicular communication, but the authors simply concluded the superiority of one scheme over others based on the number of missed detected packets only, assuming the same threshold for all the schemes under consideration. However, the consideration of false-alarm probability exposes a fundamental flaw of this approach. In essence, a packet detection scheme can trade off between the probabilities of false alarm and missed detection by adjusting its threshold. Lowering the detection threshold reduces the probability of missed detection while simultaneously increasing the probability of false alarm. Therefore, simply comparing missed detections without considering false alarms, or comparing false alarms without considering missed detections, is not reasonable.

\begin{figure}[htbp]
  \centering
  \includegraphics[width=0.45\textwidth]{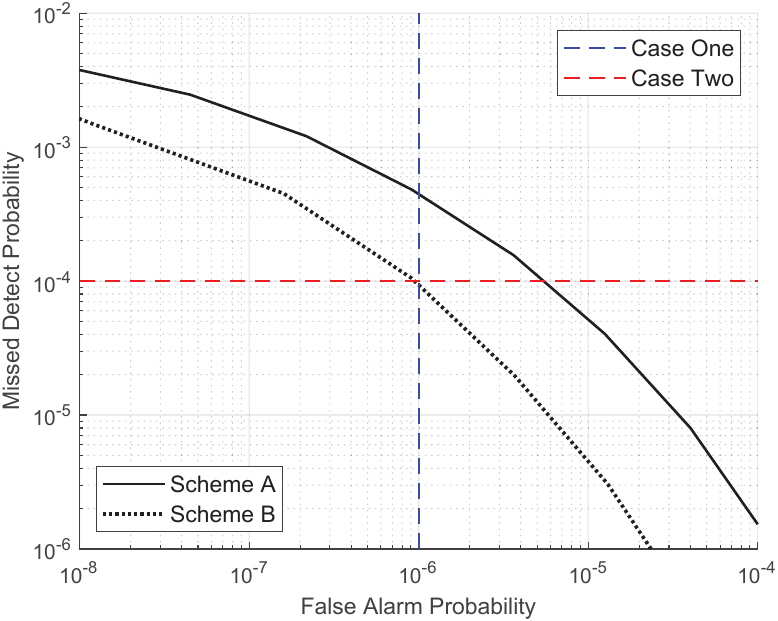}\\
  \captionsetup{font={small}}
  \caption{Two example MD-FA curves for illustrating our ``Pareto comparison" benchmarking method.}
\label{fig:7}
\end{figure}

We propose rigorous benchmarking by ``Pareto comparison". Suppose that we have two packet detection schemes, A and B. In general, we can adjust ${\rho _A}$ and ${\rho _B}$ to obtain the tradeoff curves for the operating points $\left( {P_{FA}^A({\rho _A}),P_{MD}^A({\rho _A})} \right)$ and $\left( {P_{FA}^B({\rho _B}),P_{MD}^B({\rho _B})} \right)$, respectively. To illustrate our point, in Fig. \ref{fig:7}, we plot an example of $\left( {P_{FA}^A({\rho _A}),P_{MD}^A({\rho _A})} \right)$ curve and an example of $\left( {P_{FA}^B({\rho _B}),P_{MD}^B({\rho _B})} \right)$ curve for two fictitious schemes A and B. We now explain how we benchmark schemes A and B.

We can find the thresholds for schemes A and B, ${\rho _A}$ and ${\rho _B}$, such that their false-alarm probabilities are equal. For example, in Case one of Fig. \ref{fig:7}, we fix $P_{FA}^A({\rho _A}) = P_{FA}^B({\rho _B}) = {10^{ - 6}}$. Note that $\rho_A$ and $\rho_B$ are not necessarily equal for the same false-alarm probability in both schemes. For Case one, we have $P_{MD}^B({\rho _B}) < P_{MD}^A({\rho _A})$, and thus we say scheme B is superior to scheme A for this particular operating point. Alternatively, for Case two in the figure, we fix $P_{MD}^A({\rho _A}) = P_{MD}^B({\rho _B}) = {10^{ - 4}}$ and observe that $P_{FA}^B({\rho _B}) < P_{FA}^A({\rho _A})$, and thus again we say scheme B is superior to scheme A for this particular operating point. In general, in Fig. \ref{fig:7}, scheme B is superior to scheme A in the Pareto-sense in that the overall ${P_{MD}}$ versus ${P_{FA}}$ curve (referred to as the MD-FA curve) of scheme B is lower than that of scheme A.

\begin{figure}[htbp]
  \centering
  \subfloat[]  {\label{fig:8} \includegraphics[width=0.45\textwidth]{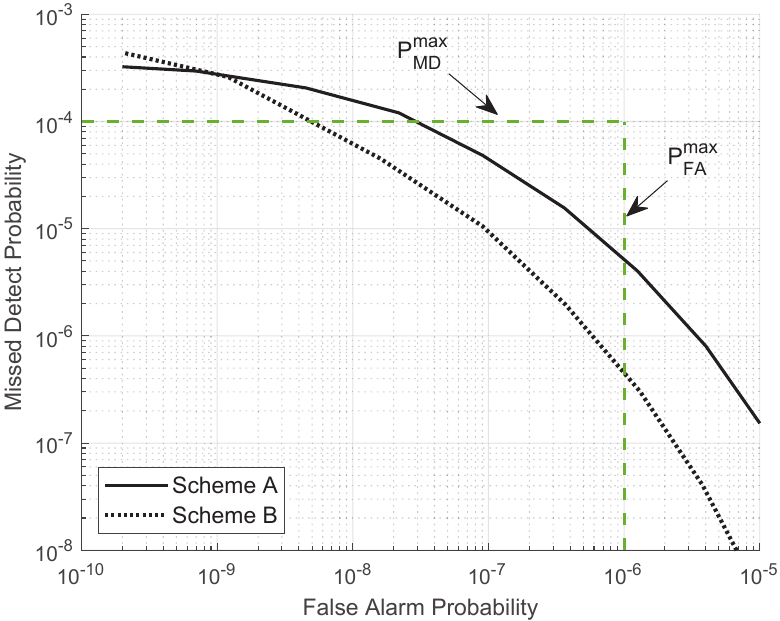}}
  \quad \quad
  \subfloat[]  {\label{fig:9} \includegraphics[width=0.45\textwidth]{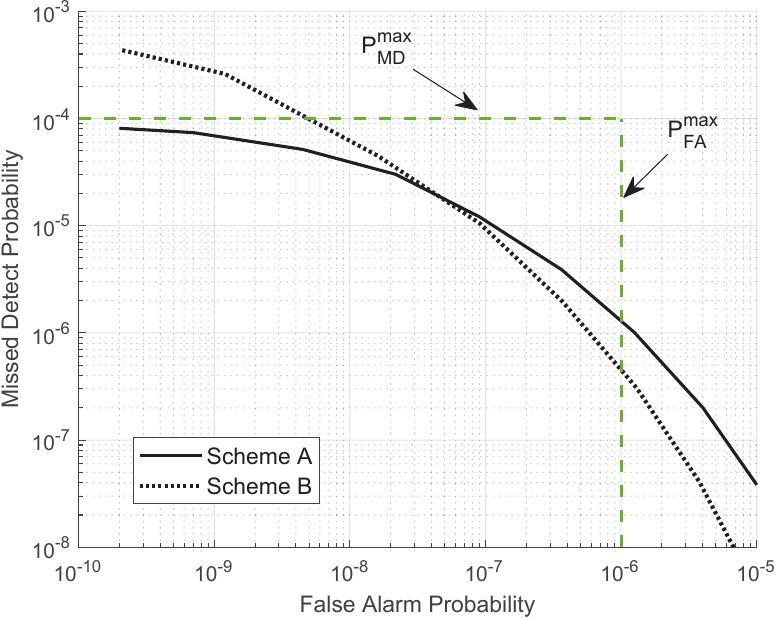}}
  \captionsetup{font={small}}
  \caption{(a) An example to illustrate that one scheme is superior to the other within a specific region of interest. (b) An example to illustrate that one scheme is comparable with the other within a specific region of interest.}
\end{figure}

If the two curves crisscross each other, it is inconclusive as to which scheme is superior. However, as illustrated in Fig. \ref{fig:8}, if the curve of scheme B is consistently lower than that of scheme A within a specific region of interest (e.g., false-alarm probability not exceeding ${10^{ - 6}}$ and missed-detection probability not exceeding ${10^{ - 4}}$), we can conclude that scheme B outperforms scheme A in that particular region (although the two curves may still crisscross outside the region of interest). Conversely, if the two curves intersect within the region of interest (as illustrated in Fig. \ref{fig:9}), we consider scheme A and scheme B to be comparable in that region, resulting in a ``draw" in terms of benchmarking the two schemes.

\begin{figure*}[htbp]
  \centering
  \includegraphics[width=0.9\textwidth]{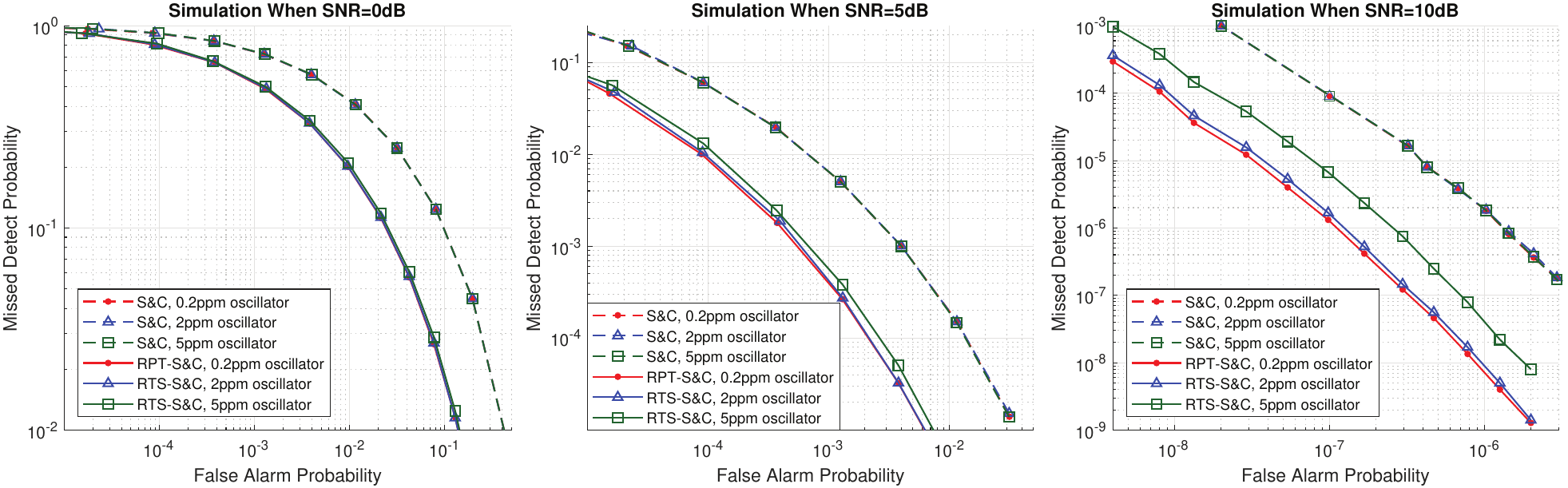}\\
  \captionsetup{font={small}}
  \caption{RP-S\&C versus conventional S\&C under various CFO and SNR conditions.}
\label{fig:10}
\end{figure*}

Given the above context, we now examine the performance of RP-S\&C and conventional S\&C. We assume a 0.2ppm/2ppm/5ppm oscillator offset in accordance with the state-of-the-art/typical/worst CFO condition that one may encounter in modern communication hardware.\footnote{We conducted real-world experiments to test the CFOs of several commercial WiFi devices. Additionally, we examined the CFOs of a well-known open-source wireless channel dataset \cite{ref19}. Our experiments revealed that the oscillator offsets of the tested hardware and the evaluated dataset are limited to a maximum of 2ppm. Hence, we consider 2ppm to be the typical oscillator offset. Moreover, we reviewed state-of-the-art research efforts published in top semiconductor journals/conferences \cite{ref20,ref21}. We found that the oscillator offsets reported in these studies do not exceed 0.2ppm. Consequently, we consider 0.2ppm to be the state-of-the-art oscillator offset. For extreme cases, we assume a 5ppm oscillator offset as the worst-case scenario.} Fig. \ref{fig:10} shows that the MD-FA curves of RP-S\&C consistently lie below those of conventional S\&C in various practical SNR and CFO settings, validating our statement in Section \ref{sec-II}-A that taking the real part of ${a}[n]$ is advantageous for reliable packet detections.

\section{Multi-antenna Packet Detection: Analysis and Optimizations}\label{sec-IV}
\subsection{Problem Formulation}
Assume that there are ${N_R}$ antennas in a receiver. Let us denote the ${a_R}[n]$, $b[n]$, and $r[n]$ of antenna $j$ by ${a_{R,j}}[n]$, ${b_j}[n]$, and ${r_j}[n]$, respectively. We want to combine ${r_j}[n]$ with carefully chosen weights $w_j$ so that the post-combined $r[n]$ yields good packet-detection performance in terms of false alarm or missed detection (or both). Let ${r^M}[n]$ represent the post-combined $r[n]$ in the multi-antenna case to distinguish it from $r[n]$ in the single-antenna case. The general expression of $r^M[n]$ is
\begin{equation}\label{eq030}
{r^M}[n] = \sum\limits_{j = 1}^{{N_R}} {{w_j}{r_j}[n]}  = \sum\limits_{j = 1}^{{N_R}} {{w_j}\left( {{a_{R,j}}[n] - \rho {b_j}[n]} \right)}
\end{equation}

The rest of this section investigates the optimal assignment for weight vector ${\bf{w}} = \left\{ {{w_1},...,{w_{{N_R}}}} \right\}$. Recall from the discussion in Section \ref{sec-II} that both false alarm and missed detection are important aspects of a packet detection algorithm. For a given threshold $\rho$, the weight vector minimizing false-alarm probability is different from that minimizing missed-detection probability.

{\bf{a. Minimizing False-Alarm Probability}}

Assume that there is no packet, as in (\ref{eq026}), ${r_{N,j}}[n]$ can be approximated as a Gaussian random variable:
\begin{equation}\label{eq031}
{r_{N,j}}[n] \sim N\left( { - \rho {\kern 1pt} {\kern 1pt} {\kern 1pt} {\kern 1pt} ,{\kern 1pt} {\kern 1pt} {\kern 1pt} {\kern 1pt} {\kern 1pt} {\kern 1pt} {\kern 1pt} \frac{{1 + {\rho ^2}}}{{2\eta }}} \right)
\end{equation}

Thus, we have
\begin{equation}\label{eq032}
E\left( {r_N^M[n]} \right) = E\left( {\sum\limits_{j = 1}^{{N_R}} {{w_j}{r_{N,j}}[n]} } \right) = \sum\limits_{j = 1}^{{N_R}} {{w_j}E\left( {{r_{N,j}}[n]} \right)}  =  - \rho  \cdot \sum\limits_{j = 1}^{{N_R}} {{w_j}}
\end{equation}
and
\begin{equation}\label{eq033}
Var\left( {r_N^M[n]} \right) = Var\left( {\sum\limits_{j = 1}^{{N_R}} {{w_j}{r_{N,j}}[n]} } \right) = \sum\limits_{j = 1}^{{N_R}} {w_j^2 \cdot Var\left( {{r_{N,j}}[n]} \right)}  = \frac{{1 + {\rho ^2}}}{{2\eta }}\sum\limits_{j = 1}^{{N_R}} {w_j^2}
\end{equation}

As in (\ref{eq029}), the false-alarm probability in the multi-antenna case is given by
\begin{equation}\label{eq034}
P_{FA}^M = Q\left( { - \frac{{E\left( {{r_N}[n]} \right)}}{{\sqrt {Var\left( {{r_N}[n]} \right)} }}} \right) = Q\left( {\sqrt {\frac{{2\eta }}{{1 + {\rho ^2}}}}  \cdot {{\rho  \cdot \sum\limits_{j = 1}^{{N_R}} {{w_j}} } \mathord{\left/
 {\vphantom {{\rho  \cdot \sum\limits_{j = 1}^{{N_R}} {{w_j}} } {\sqrt {\sum\limits_{j = 1}^{{N_R}} {w_j^2} } }}} \right.
 \kern-\nulldelimiterspace} {\sqrt {\sum\limits_{j = 1}^{{N_R}} {w_j^2} } }}} \right)
\end{equation}

We note from (\ref{eq034}) that, for any weight vector $w$, the weight vector scaled by a constant $c > 0$ yields the same $P_{FA}^M$. We can impose a normalization condition $\sum\nolimits_{j = 1}^{{N_R}} {{w_j}}  = 1$ without changing the outcome. Hence, we can formulate the optimization problem as
\begin{equation}\label{eq035}
\max f({\bf{w}}) = {\left( {\sum\limits_{j = 1}^{{N_R}} {w_j^2} } \right)^{ - 1}},{\kern 1pt} {\kern 1pt} {\kern 1pt} {\kern 1pt} {\kern 1pt} {\kern 1pt} {\text{subject}}{\kern 1pt} {\kern 1pt} {\text{ to}} {\kern 1pt} {\kern 1pt} \sum\limits_{j = 1}^{{N_R}} {{w_j}}  = 1,{\kern 1pt} {\kern 1pt} {\kern 1pt} {\kern 1pt} {\rm{and}}{\kern 1pt} {\kern 1pt} {\kern 1pt} {\kern 1pt} {w_j} \ge 0{\kern 1pt} {\kern 1pt} {\kern 1pt} {\kern 1pt} {\kern 1pt} \forall j \in \left\{ {1,...,{N_R}} \right\}
\end{equation}

{\bf{b. Minimizing Missed-Detection Probability}}

Assume that there is a packet, as in (\ref{eq025}), ${r_{P,j}}[n]$ can be approximated as a Gaussian random variable:
\begin{equation}\label{eq036}
{r_{P,j}}[n] \sim N\left( {(1 - \rho ){\gamma _j} - \rho {\kern 1pt} {\kern 1pt} {\kern 1pt} {\kern 1pt} ,{\kern 1pt} {\kern 1pt} {\kern 1pt} {\kern 1pt} {\kern 1pt} {\kern 1pt} {\kern 1pt} \frac{{{{(1 - \rho )}^2}}}{\eta }{\gamma _j} + \frac{{1 + {\rho ^2}}}{{2\eta }}} \right)
\end{equation}

Thus, we have
\begin{equation}\label{eq037}
E\left( {r_P^M[n]} \right) = E\left( {\sum\limits_{j = 1}^{{N_R}} {{w_j}{r_{P,j}}[n]} } \right) = \sum\limits_{j = 1}^{{N_R}} {{w_j}E\left( {{r_{P,j}}[n]} \right)}  = \sum\limits_{j = 1}^{{N_R}} {{w_j}\left[ {(1 - \rho ){\gamma _j} - \rho } \right]}
\end{equation}
and
\begin{equation}\label{eq038}
\begin{array}{l}
Var\left( {r_P^M[n]} \right) = Var\left( {\sum\limits_{j = 1}^{{N_R}} {{w_j}\cdot {r_{P,j}}[n]} } \right) = \sum\limits_{j = 1}^{{N_R}} {w_j^2  \cdot Var\left( {{r_{P,j}}[n]} \right)}  = \sum\limits_{j = 1}^{{N_R}} {w_j^2 \left[ {\frac{{{{(1 - \rho )}^2}}}{\eta }{\gamma _j} + \frac{{1 + {\rho ^2}}}{{2\eta }}} \right]}
\end{array}
\end{equation}

As in (\ref{eq028}), the missed-detection probability is given by
\begin{equation}\label{eq039}
P_{MD}^M = Q\left( {\frac{{E\left( {r_P^M[n]} \right)}}{{\sqrt {Var\left( {r_P^M[n]} \right)} }}} \right) = Q\left( {\sqrt \eta  \frac{{\sum\limits_{j = 1}^{{N_R}} {{w_j}\left[ {(1 - \rho ){\gamma _j} - \rho } \right]} {\kern 1pt} }}{{\sqrt {\sum\limits_{j = 1}^{{N_R}} {w_j^2\left[ {{{(1 - \rho )}^2}{\gamma _j} + {{\left( {1 + {\rho ^2}} \right)} \mathord{\left/
 {\vphantom {{\left( {1 + {\rho ^2}} \right)} 2}} \right.
 \kern-\nulldelimiterspace} 2}} \right]} } }}} \right)
\end{equation}

From (\ref{eq039}), we can therefore formulate the optimization problem as
\begin{equation}\label{eq040}
\begin{array}{l}
\max g({\bf{w}}) = {{{{\left( {\sum\limits_{j = 1}^{{N_R}} {{w_j}\left[ {(1 - \rho ){\gamma _j} - \rho } \right]} } \right)}^2}} \mathord{\left/
 {\vphantom {{{{\left( {\sum\limits_{j = 1}^{{N_R}} {{w_j}\left[ {(1 - \rho ){\gamma _j} - \rho } \right]} } \right)}^2}} {\sum\limits_{j = 1}^{{N_R}} {w_j^2\left[ {{{(1 - \rho )}^2}{\gamma _j} + {{\left( {1 + {\rho ^2}} \right)} \mathord{\left/
 {\vphantom {{\left( {1 + {\rho ^2}} \right)} 2}} \right.
 \kern-\nulldelimiterspace} 2}} \right]} }}} \right.
 \kern-\nulldelimiterspace} {\sum\limits_{j = 1}^{{N_R}} {w_j^2\left[ {{{(1 - \rho )}^2}{\gamma _j} + {{\left( {1 + {\rho ^2}} \right)} \mathord{\left/
 {\vphantom {{\left( {1 + {\rho ^2}} \right)} 2}} \right.
 \kern-\nulldelimiterspace} 2}} \right]} }}{\kern 1pt} {\kern 1pt} {\kern 1pt} {\kern 1pt} {\kern 1pt} {\kern 1pt} {\kern 1pt} {\kern 1pt} {\kern 1pt} {\kern 1pt} {\kern 1pt} \\
{\kern 1pt} {\text{subject  to  }}{\kern 1pt} \sum\limits_{j = 1}^{{N_R}} {{w_j}}  = 1,{\kern 1pt} {\kern 1pt} {\kern 1pt} {\kern 1pt} {\rm{and}}{\kern 1pt} {\kern 1pt} {\kern 1pt} {\kern 1pt} {w_j} \ge 0{\kern 1pt} {\kern 1pt} {\kern 1pt} {\kern 1pt} \forall j \in \left\{ {1,...,{N_R}} \right\}
\end{array}
\end{equation}

Note from (\ref{eq035}) and (\ref{eq040}) that both minimizing false alarm and minimizing missed detection are subjected to the constraint of $\sum\nolimits_{j = 1}^{{N_R}} {{w_j}}  = 1$ and ${w_j} \ge 0{\kern 1pt} {\kern 1pt} {\kern 1pt} {\kern 1pt} \forall j \in \left\{ {1,...,{N_R}} \right\}$. In the rest of this paper, we call ${\bf{w}} = \left\{ {{w_1},...,{w_{{N_R}}}} \right\}$ a feasible weight vector only if it satisfies the constraint.

\subsection{Optimal Weights for False Alarm (WFA) and Missed Detection (WMD)}
\begin{proposition}\label{proposition:1}
For a receiver with $N_R$ antennas, the equal-weight assignment ${w_j} = {1 \mathord{\left/ {\vphantom {1 {{N_R}}}} \right. \kern-\nulldelimiterspace} {{N_R}}}$ to ${r_{N,j}}[n]$ yields the minimum $P_{FA}^M$.
\end{proposition}
\begin{NewProof}
\begin{equation}\label{eq041}
\mathop {\arg \max }\limits_{\scriptstyle{\kern 1pt} {\kern 1pt}{\kern 1pt} {\kern 1pt}{\kern 1pt} {\kern 1pt}{\kern 1pt} {\kern 1pt}{\kern 1pt} {\kern 1pt} {\kern 1pt} {\kern 1pt} {\kern 1pt} {\kern 1pt} {\kern 1pt} {\kern 1pt} {\kern 1pt} {\kern 1pt} {\kern 1pt} {\kern 1pt} {\kern 1pt} {\kern 1pt} {\kern 1pt} {\kern 1pt} {\kern 1pt} {\kern 1pt} {\kern 1pt} {\kern 1pt} {\kern 1pt} {\kern 1pt} {\kern 1pt} {\kern 1pt} {\bf{w}}\hfill\atop
\scriptstyle s.t.\sum {{w_j}}  = 1,{w_j} \ge 0\hfill} f({\bf{w}}) = \mathop {\arg \min }\limits_{\scriptstyle{\kern 1pt} {\kern 1pt} {\kern 1pt} {\kern 1pt} {\kern 1pt} {\kern 1pt} {\kern 1pt} {\kern 1pt} {\kern 1pt} {\kern 1pt} {\kern 1pt} {\kern 1pt} {\kern 1pt}{\kern 1pt} {\kern 1pt}{\kern 1pt} {\kern 1pt}{\kern 1pt} {\kern 1pt}{\kern 1pt} {\kern 1pt} {\kern 1pt} {\kern 1pt} {\kern 1pt} {\kern 1pt} {\kern 1pt} {\kern 1pt} {\kern 1pt} {\kern 1pt} {\kern 1pt} {\kern 1pt} {\kern 1pt} {\bf{w}}\hfill\atop
\scriptstyle s.t.\sum {{w_j}}  = 1,{w_j} \ge 0\hfill} \sum\limits_{j = 1}^{{N_R}} {w_j^2}
\end{equation}

It is easy to see that the answer to (\ref{eq041}) is found by setting ${w_j} = {1 \mathord{\left/ {\vphantom {1 {{N_R}}}} \right. \kern-\nulldelimiterspace} {{N_R}}}$ for all $j$.
\end{NewProof}

Finding the optimal weights for missed-detection probability is more challenging. Let us look at the derivative of $g({\bf{w}})$ over ${w_j}$:
\begin{equation}\label{eq042}
\frac{{\partial g({\bf{w}})}}{{\partial {w_j}}} \buildrel \Delta \over = \frac{{n({w_j})}}{{d({w_j})}}
\end{equation}
where the denominator $d({w_j})$ is always positive, i.e.,
\begin{equation}\label{eq043}
d({w_j}) = {\left( {\sum\limits_{m = 1}^{{N_R}} {w_m^2\left[ {{{(1 - \rho )}^2}{\gamma _m} + \frac{{1 + {\rho ^2}}}{2}} \right]} } \right)^2} > 0
\end{equation}
and the numerator $n({w_j})$ is
\begin{equation}\label{eq044}
\small
\begin{array}{l}
n({w_j}) = \left( {\sum\limits_{m = 1}^{{N_R}} {w_m^2\left[ {{{(1 - \rho )}^2}{\gamma _m} + \frac{{1 + {\rho ^2}}}{2}} \right]} } \right) \cdot 2\left( {\sum\limits_{m = 1}^{{N_R}} {{w_m}\left[ {(1 - \rho ){\gamma _m} - \rho } \right]} } \right) \cdot \left[ {(1 - \rho ){\gamma _j} - \rho } \right]\\
\;\;\;\;\;\;\;\;\;\;\;\;\;\;\;\;\;\;\;\;\;\;\;\;\;\;\;\;\;\;\;\;\;\;\;\;\;\;\;\;\;\;\;\;\;\;\;\;\;\;\;\;\;\;\;\; - {\left( {\sum\limits_{m = 1}^{{N_R}} {{w_m}\left[ {(1 - \rho ){\gamma _m} - \rho } \right]} } \right)^2} \cdot 2{w_j}\left[ {{{(1 - \rho )}^2}{\gamma _j} + \frac{{1 + {\rho ^2}}}{2}} \right]\\
{\kern 1pt} {\kern 1pt} {\kern 1pt} {\kern 1pt} {\kern 1pt} {\kern 1pt} {\kern 1pt} {\kern 1pt} {\kern 1pt}  = \underbrace {2\left( {\sum\limits_{m = 1}^{{N_R}} {{w_m}\left[ {(1 - \rho ){\gamma _m} - \rho } \right]} } \right)}_{Term{\kern 1pt} {\kern 1pt} {\kern 1pt} A}\underbrace {\left\{ \begin{array}{l}
\left( {\left[ {(1 - \rho ){\gamma _j} - \rho } \right] \cdot \sum\limits_{m = 1}^{{N_R}} {w_m^2\left[ {{{(1 - \rho )}^2}{\gamma _m} + \frac{{1 + {\rho ^2}}}{2}} \right]} } \right)\\
\; - \left[ {{{(1 - \rho )}^2}{\gamma _j} + \frac{{1 + {\rho ^2}}}{2}} \right] \cdot \left( {\sum\limits_{m = 1}^{{N_R}} {{w_m}{w_j}\left[ {(1 - \rho ){\gamma _m} - \rho } \right]} } \right)
\end{array} \right\}}_{Term{\kern 1pt} {\kern 1pt} {\kern 1pt} B}\\
{\kern 1pt} {\kern 1pt} {\kern 1pt} {\kern 1pt} {\kern 1pt} {\kern 1pt} {\kern 1pt} {\kern 1pt} {\kern 1pt}  = \underbrace {2\left( {\sum\limits_{m = 1}^{{N_R}} {{w_m}\left[ {(1 - \rho ){\gamma _m} - \rho } \right]} } \right)}_{Term{\kern 1pt} {\kern 1pt} {\kern 1pt} A}\underbrace {\left\{ {\sum\limits_{m = 1}^{{N_R}} {{w_m}\left[ \begin{array}{l}
({w_m} - {w_j})\left( {{{(1 - \rho )}^3}{\gamma _m}{\gamma _j} - \frac{{(1 + {\rho ^2})\rho }}{2}} \right)\\
\; + ({w_m}{\gamma _j} - {w_j}{\gamma _m})\frac{{(1 + {\rho ^2})(1 - \rho )}}{2} + ({w_j}{\gamma _j} - {w_m}{\gamma _m})\rho {(1 - \rho )^2}
\end{array} \right]} } \right\}}_{Term{\kern 1pt} {\kern 1pt} {\kern 1pt} B}\\
{\kern 1pt} {\kern 1pt} {\kern 1pt} {\kern 1pt} {\kern 1pt} {\kern 1pt} {\kern 1pt} {\kern 1pt}  = \underbrace {2\left( {\sum\limits_{m = 1}^{{N_R}} {{w_m}\left[ {(1 - \rho ){\gamma _m} - \rho } \right]} } \right)}_{Term{\kern 1pt} {\kern 1pt} {\kern 1pt} A}\underbrace {\left\{ \begin{array}{l}
\left[ {(1 - \rho ){\gamma _j} - \rho } \right] \cdot \sum\limits_{m \ne j} {w_m^2\left[ {{{(1 - \rho )}^2}{\gamma _m} + \frac{{1 + {\rho ^2}}}{2}} \right]} \\
\; - {w_j}\left[ {{{(1 - \rho )}^2}{\gamma _j} + \frac{{1 + {\rho ^2}}}{2}} \right] \cdot \sum\limits_{m \ne j} {{w_m}\left[ {(1 - \rho ){\gamma _m} - \rho } \right]}
\end{array} \right\}}_{Term{\kern 1pt} {\kern 1pt} {\kern 1pt} B}
\end{array}
\normalsize
\end{equation}

In (\ref{eq044}), we write $n({w_j})$ as the product of term A and term B. In term A, we can impose a practical constraint\footnote{We will later show in Section V that a typical $\rho$ in practical multi-antenna systems is no larger than 0.5, which means ${\rho  \mathord{\left/ {\vphantom {\rho  {(1 - \rho )}}} \right. \kern-\nulldelimiterspace} {(1 - \rho )}}$ is no larger than 0dB. If an antenna $j$  has SNR ${\gamma _j}$ of 0 dB or lower, it will not contribute much to packet detection and packet decoding, and we might as well omit it in both considerations. In other words, in this analysis, we assume that antennas with SNR of less than ${\rho  \mathord{\left/ {\vphantom {\rho  {(1 - \rho )}}} \right. \kern-\nulldelimiterspace} {(1 - \rho )}}$ would not be used for packet detection purposes.} of ${\gamma _m} > {\rho  \mathord{\left/ {\vphantom {\rho  {(1 - \rho )}}} \right. \kern-\nulldelimiterspace} {(1 - \rho )}}$ for every antenna so that we have $(1 - \rho ){\gamma _m} - \rho  > 0,{\kern 1pt} {\kern 1pt} {\kern 1pt} {\kern 1pt} {\kern 1pt} {\kern 1pt} \forall m \in \left\{ {1,...{N_R}} \right\}$. With $\sum\nolimits_{j = 1}^{{N_R}} {{w_j} = 1} ,{\kern 1pt} {\kern 1pt} {\kern 1pt} {\kern 1pt} {w_j} \ge 0$ and $(1 - \rho ){\gamma _m} - \rho  > 0$, we know that term A is positive. In term B, the terms in which $m = j$ in the two summations cancel out each other, so we exclude them in the summations and obtain the final form of term B in the last line.

We note that for a locally optimal solution, we require $\frac{{\partial g({\bf{w}})}}{{\partial {w_j}}} = \frac{{n({w_j})}}{{d({w_j})}} = 0$ for all $j \in \left\{ {1,...,{N_R}} \right\}$. Thus, term B should be zero and the following equation should hold:
\begin{equation}\label{eq045}
\small
\begin{array}{l}
\left[ {(1 - \rho ){\gamma _j} - \rho } \right] \cdot \sum\limits_{m \ne j} {w_m^2\left[ {{{(1 - \rho )}^2}{\gamma _m} + \frac{{1 + {\rho ^2}}}{2}} \right]}  = {w_j}\left[ {{{(1 - \rho )}^2}{\gamma _j} + \frac{{1 + {\rho ^2}}}{2}} \right] \cdot \sum\limits_{m \ne j} {{w_m}\left[ {(1 - \rho ){\gamma _m} - \rho } \right]}
\end{array}
\normalsize
\end{equation}

From (\ref{eq045}), we have
\begin{equation}\label{eq046}
{w_j} = \frac{{\sum\limits_{m \ne j} {w_m^2\left[ {{{(1 - \rho )}^2}{\gamma _m} + {{(1 + {\rho ^2})} \mathord{\left/
 {\vphantom {{(1 + {\rho ^2})} 2}} \right.
 \kern-\nulldelimiterspace} 2}} \right]\left[ {(1 - \rho ){\gamma _j} - \rho } \right]} }}{{\sum\limits_{m \ne j} {{w_m}\left[ {{{(1 - \rho )}^2}{\gamma _j} + {{(1 + {\rho ^2})} \mathord{\left/
 {\vphantom {{(1 + {\rho ^2})} 2}} \right.
 \kern-\nulldelimiterspace} 2}} \right]\left[ {(1 - \rho ){\gamma _m} - \rho } \right]} }},\,{\kern 1pt} {\kern 1pt} {\kern 1pt} {\kern 1pt} {\kern 1pt} {\kern 1pt} {\kern 1pt} {\kern 1pt} {\kern 1pt} j \in \left\{ {1,...,{N_R}} \right\}
\end{equation}

To investigate (\ref{eq046}), we start from the simple two-antenna case, i.e., we only have antenna 1 and antenna 2. We have
\begin{equation}\label{eq047}
{w_1} = \frac{{w_2^2[{{(1 - \rho )}^2}{\gamma _2} + {{(1 + {\rho ^2})} \mathord{\left/
 {\vphantom {{(1 + {\rho ^2})} 2}} \right.
 \kern-\nulldelimiterspace} 2}][(1 - \rho ){\gamma _1} - \rho ]}}{{{w_2}[{{(1 - \rho )}^2}{\gamma _1} + {{(1 + {\rho ^2})} \mathord{\left/
 {\vphantom {{(1 + {\rho ^2})} 2}} \right.
 \kern-\nulldelimiterspace} 2}][(1 - \rho ){\gamma _2} - \rho ]}}
\end{equation}
and
\begin{equation}\label{ref048}
{w_2} = \frac{{w_1^2[{{(1 - \rho )}^2}{\gamma _1} + {{(1 + {\rho ^2})} \mathord{\left/
 {\vphantom {{(1 + {\rho ^2})} 2}} \right.
 \kern-\nulldelimiterspace} 2}][(1 - \rho ){\gamma _2} - \rho ]}}{{{w_1}[{{(1 - \rho )}^2}{\gamma _2} + {{(1 + {\rho ^2})} \mathord{\left/
 {\vphantom {{(1 + {\rho ^2})} 2}} \right.
 \kern-\nulldelimiterspace} 2}][(1 - \rho ){\gamma _1} - \rho ]}}
\end{equation}

The above gives
\begin{equation}\label{eq049}
\begin{array}{l}
\frac{{{w_1}}}{{{w_2}}} = \frac{{[{{(1 - \rho )}^2}{\gamma _2} + {{(1 + {\rho ^2})} \mathord{\left/
 {\vphantom {{(1 + {\rho ^2})} 2}} \right.
 \kern-\nulldelimiterspace} 2}][(1 - \rho ){\gamma _1} - \rho ]}}{{[{{(1 - \rho )}^2}{\gamma _1} + {{(1 + {\rho ^2})} \mathord{\left/
 {\vphantom {{(1 + {\rho ^2})} 2}} \right.
 \kern-\nulldelimiterspace} 2}][(1 - \rho ){\gamma _2} - \rho ]}} = {{\frac{{[(1 - \rho ){\gamma _1} - \rho ]}}{{[{{(1 - \rho )}^2}{\gamma _1} + {{(1 + {\rho ^2})} \mathord{\left/
 {\vphantom {{(1 + {\rho ^2})} 2}} \right.
 \kern-\nulldelimiterspace} 2}]}}} \mathord{\left/
 {\vphantom {{\frac{{[(1 - \rho ){\gamma _1} - \rho ]}}{{[{{(1 - \rho )}^2}{\gamma _1} + {{(1 + {\rho ^2})} \mathord{\left/
 {\vphantom {{(1 + {\rho ^2})} 2}} \right.
 \kern-\nulldelimiterspace} 2}]}}} {\frac{{[(1 - \rho ){\gamma _2} - \rho ]}}{{[{{(1 - \rho )}^2}{\gamma _2} + {{(1 + {\rho ^2})} \mathord{\left/
 {\vphantom {{(1 + {\rho ^2})} 2}} \right.
 \kern-\nulldelimiterspace} 2}]}}}}} \right.
 \kern-\nulldelimiterspace} {\frac{{[(1 - \rho ){\gamma _2} - \rho ]}}{{[{{(1 - \rho )}^2}{\gamma _2} + {{(1 + {\rho ^2})} \mathord{\left/
 {\vphantom {{(1 + {\rho ^2})} 2}} \right.
 \kern-\nulldelimiterspace} 2}]}}}}
 \end{array}
\end{equation}

As in Subsection B, we impose the constraint of $\sum\nolimits_{j = 1}^{{N_R}} {{w_j}}  = 1$. Thus, a feasible locally optimal solution for the two-antenna case is given by
\begin{equation}\label{eq050}
{w_j} = c\frac{{[(1 - \rho ){\gamma _j} - \rho ]}}{{[{{(1 - \rho )}^2}{\gamma _j} + {{(1 + {\rho ^2})} \mathord{\left/
 {\vphantom {{(1 + {\rho ^2})} 2}} \right.
 \kern-\nulldelimiterspace} 2}]}},{\kern 1pt} {\kern 1pt} {\kern 1pt} {\kern 1pt} {\kern 1pt} {\kern 1pt} {\kern 1pt} {\kern 1pt} j = 1,2,{\kern 1pt} {\kern 1pt} {\kern 1pt} {\kern 1pt} {\kern 1pt} {\kern 1pt} {\kern 1pt} {\kern 1pt} c = {\left( {\sum\limits_{m = 1}^2 {\frac{{[(1 - \rho ){\gamma _m} - \rho ]}}{{[{{(1 - \rho )}^2}{\gamma _m} + {{(1 + {\rho ^2})} \mathord{\left/
 {\vphantom {{(1 + {\rho ^2})} 2}} \right.
 \kern-\nulldelimiterspace} 2}]}}} } \right)^{ - 1}}
\end{equation}

We now extend our analysis to cases with more than two antennas. We shall see that the solution form of (\ref{eq050}) is retained for the general case. With the general expression of ${w_j}$ given in (\ref{eq046}), we can verify that a feasible locally optimal solution for a $N_R$-antenna case is
\begin{equation}\label{eq051}
{w_j} = c\frac{{(1 - \rho ){\gamma _j} - \rho }}{{{{(1 - \rho )}^2}{\gamma _j} + {{(1 + {\rho ^2})} \mathord{\left/
 {\vphantom {{(1 + {\rho ^2})} 2}} \right.
 \kern-\nulldelimiterspace} 2}}},{\kern 1pt} {\kern 1pt} {\kern 1pt} {\kern 1pt} {\kern 1pt} {\kern 1pt} j = 1,2,...{N_R},{\kern 1pt} {\kern 1pt} {\kern 1pt} {\kern 1pt} {\kern 1pt} {\kern 1pt} {\kern 1pt} {\kern 1pt} {\kern 1pt} c = {\left( {\sum\limits_{m = 1}^{{N_R}} {\frac{{[(1 - \rho ){\gamma _m} - \rho ]}}{{[{{(1 - \rho )}^2}{\gamma _m} + {{(1 + {\rho ^2})} \mathord{\left/
 {\vphantom {{(1 + {\rho ^2})} 2}} \right.
 \kern-\nulldelimiterspace} 2}]}}} } \right)^{ - 1}}
\end{equation}

In the rest of this paper, we denote the weight vector calculated according to (\ref{eq051}) by ${{\bf{w}}^o}$. We now prove that ${{\bf{w}}^o}$ is the unique solution that yields the global maximum $g({\bf{w}})$.

\begin{proposition}\label{proposition:2}
If a feasible weight vector ${\bf{w}}$ does not satisfy (\ref{eq051}), i.e., ${\bf{w}} \ne {{\bf{w}}^o}$, then ${\bf{w}}$ is non-optimal. Thus, ${{\bf{w}}^o}$ in (\ref{eq051}) is the unique optimal solution to minimizing ${P_{MD}}$ as per (\ref{eq045}).
\end{proposition}
\begin{NewProof}
We prove that there is another solution ${\bf{w'}}$ that yields $g({\bf{w'}}) > g({\bf{w}})$.

We first note that it is not possible that ${w_j} < w_j^o$ for all $j \in \left\{ {1,...,{N_R}} \right\}$ or ${w_j} > w_j^o$ for all $j \in \left\{ {1,...,{N_R}} \right\}$ because that would mean $\sum\nolimits_{j = 1}^{{N_R}} {{w_j}}  < 1$ or $\sum\nolimits_{j = 1}^{{N_R}} {{w_j}}  > 1$. Thus, given that ${\bf{w}} \ne {{\bf{w}}^o}$ and that ${\bf{w}}$ is feasible (i.e., $\sum\nolimits_{j = 1}^{{N_R}} {{w_j}}  = 1$), there must be at least one $k$ such that ${{{w_k}} \mathord{\left/ {\vphantom {{{w_k}} {w_k^o}}} \right. \kern-\nulldelimiterspace} {w_k^o}} > 1$ and at least one $i$ such that ${{{w_i}} \mathord{\left/ {\vphantom {{{w_i}} {w_i^o}}} \right. \kern-\nulldelimiterspace} {w_i^o}} < 1$ and . Let us refer to ${{{w_j}} \mathord{\left/ {\vphantom {{{w_j}} {w_j^o}}} \right.  \kern-\nulldelimiterspace} {w_j^o}}$ as the weight ratio of index $j$. In general, there could be multiple weight ratios of different indexes that attain the maximum, and multiple weight ratios of different indexes that attain the minimum. Let the respective sets be

\begin{equation}\label{eq052}
{\bf{K}} = \left\{ {k:\frac{{{w_k}}}{{w_k^o}} = {{\max }_j}\frac{{{w_j}}}{{w_j^o}}} \right\}{\kern 1pt} {\kern 1pt} {\kern 1pt} {\kern 1pt} {\kern 1pt} {\kern 1pt} {\kern 1pt} {\kern 1pt} {\kern 1pt} {\kern 1pt} {\kern 1pt} {\kern 1pt} {\kern 1pt} {\kern 1pt} {\kern 1pt} {\kern 1pt} {\kern 1pt} {\kern 1pt} {\kern 1pt} {\kern 1pt} {\kern 1pt} {\kern 1pt} {\kern 1pt} {\kern 1pt} {\kern 1pt} {\kern 1pt} {\kern 1pt} {\kern 1pt} {\kern 1pt} {\kern 1pt} {\kern 1pt} {\kern 1pt} {\kern 1pt} {\kern 1pt} {\kern 1pt} {\kern 1pt} {\kern 1pt} {\kern 1pt} {\kern 1pt} {\kern 1pt} {\kern 1pt} {\kern 1pt} {\rm{and}}{\kern 1pt} {\kern 1pt} {\kern 1pt} {\kern 1pt} {\kern 1pt} {\kern 1pt} {\kern 1pt} {\kern 1pt} {\kern 1pt} {\kern 1pt} {\kern 1pt} {\kern 1pt} {\kern 1pt} {\kern 1pt} {\kern 1pt} {\kern 1pt} {\kern 1pt} {\kern 1pt} {\kern 1pt} {\kern 1pt} {\kern 1pt} {\kern 1pt} {\kern 1pt} {\kern 1pt} {\kern 1pt} {\kern 1pt} {\kern 1pt} {\kern 1pt} {\kern 1pt} {\kern 1pt} {\kern 1pt} {\kern 1pt} {\kern 1pt} {\kern 1pt} {\kern 1pt} {\kern 1pt} {\kern 1pt} {\kern 1pt} {\kern 1pt} {\kern 1pt} {\kern 1pt} {\kern 1pt} {\bf{I}} = \left\{ {i:\frac{{{w_i}}}{{w_i^o}} = {{\min }_j}\frac{{{w_j}}}{{w_j^o}}} \right\}
\end{equation}

Now, consider a $k \in {\bf{K}}$ and an $i \in {\bf{I}}$ . We have that
\begin{equation}\label{eq053}
{w_k} > \frac{{{w_j}}}{{w_j^o}}w_k^o,{\kern 1pt} {\kern 1pt} {\kern 1pt}  \forall j \notin {\bf{K}}{\kern 1pt} {\kern 1pt} {\kern 1pt} {\kern 1pt} {\kern 1pt} {\kern 1pt} {\kern 1pt} {\kern 1pt} {\kern 1pt} {\kern 1pt} {\kern 1pt} {\kern 1pt} {\kern 1pt} {\kern 1pt} {\kern 1pt} {\kern 1pt} {\kern 1pt} {\kern 1pt} {\kern 1pt} {\kern 1pt} {\kern 1pt} {\kern 1pt} {\kern 1pt} {\kern 1pt} {\kern 1pt} {\kern 1pt} {\kern 1pt} {\kern 1pt} {\kern 1pt} {\kern 1pt} {\kern 1pt} {\kern 1pt} {\kern 1pt} {\kern 1pt} {\kern 1pt} {\kern 1pt} {\kern 1pt}  {\kern 1pt} {\kern 1pt} {\kern 1pt} {\kern 1pt} {\kern 1pt} {\kern 1pt} {\kern 1pt} {\kern 1pt} {\kern 1pt} {\kern 1pt} {\kern 1pt} {\kern 1pt} {\kern 1pt} {\kern 1pt} {\kern 1pt} {\kern 1pt} {\kern 1pt} {\kern 1pt} {\kern 1pt} {\kern 1pt} {\kern 1pt} {\kern 1pt} {\kern 1pt} {\kern 1pt} {\kern 1pt} {\kern 1pt} {\kern 1pt} {\kern 1pt} {\kern 1pt} {\kern 1pt} {\rm{and}}{\kern 1pt} {\kern 1pt} {\kern 1pt} {\kern 1pt} {\kern 1pt} {\kern 1pt} {\kern 1pt} {\kern 1pt} {\kern 1pt} {\kern 1pt} {\kern 1pt} {\kern 1pt} {\kern 1pt} {\kern 1pt} {\kern 1pt} {\kern 1pt} {\kern 1pt} {\kern 1pt} {\kern 1pt} {\kern 1pt} {\kern 1pt} {\kern 1pt} {\kern 1pt} {\kern 1pt} {\kern 1pt} {\kern 1pt} {\kern 1pt} {\kern 1pt} {\kern 1pt} {\kern 1pt} {\kern 1pt} {\kern 1pt} {\kern 1pt} {\kern 1pt} {\kern 1pt} {\kern 1pt} {\kern 1pt} {\kern 1pt} {\kern 1pt} {\kern 1pt} {\kern 1pt} {\kern 1pt} {\kern 1pt} {\kern 1pt} {\kern 1pt} {\kern 1pt} {\kern 1pt} {\kern 1pt} {\kern 1pt} {\kern 1pt} {\kern 1pt} {\kern 1pt} {\kern 1pt} {\kern 1pt} {\kern 1pt} {\kern 1pt} {\kern 1pt} {\kern 1pt} {\kern 1pt} {\kern 1pt} {\kern 1pt} {\kern 1pt} {\kern 1pt} {\kern 1pt} {\kern 1pt} {\kern 1pt} {\kern 1pt}{\kern 1pt} {\kern 1pt} {\kern 1pt}{\kern 1pt} {\kern 1pt} {\kern 1pt} {\kern 1pt}{\kern 1pt} {\kern 1pt} {\kern 1pt} {\kern 1pt} {w_i} < \frac{{{w_j}}}{{w_j^o}}w_i^o,{\kern 1pt} {\kern 1pt} {\kern 1pt} {\kern 1pt} {\kern 1pt} {\kern 1pt} \forall j \notin {\bf{I}}
\end{equation}

With (\ref{eq053}), we look back to the expression given in (\ref{eq044}). We have that
\begin{equation}\label{eq054}
\small
\begin{array}{l}
n({w_k}) = \underbrace {2\left( {\sum\limits_{m = 1}^{{N_R}} {{w_m}\left[ {(1 - \rho ){\gamma _m} - \rho } \right]} } \right)}_{Term{\kern 1pt} {\kern 1pt} {\kern 1pt} A}\underbrace {\left\{ \begin{array}{l}
\left[ {(1 - \rho ){\gamma _k} - \rho } \right] \cdot \sum\limits_{m \ne k} {w_m^2\left[ {{{(1 - \rho )}^2}{\gamma _m} + \frac{{1 + {\rho ^2}}}{2}} \right]} \\
{\kern 1pt} {\kern 1pt} {\kern 1pt} {\kern 1pt}  - {w_k}\left[ {{{(1 - \rho )}^2}{\gamma _k} + \frac{{1 + {\rho ^2}}}{2}} \right] \cdot \sum\limits_{m \ne k} {{w_m}\left[ {(1 - \rho ){\gamma _m} - \rho } \right]}
\end{array} \right\}}_{Term{\kern 1pt} {\kern 1pt} {\kern 1pt} B}\\
{\kern 1pt} {\kern 1pt} {\kern 1pt} {\kern 1pt} {\kern 1pt} {\kern 1pt} {\kern 1pt} {\kern 1pt} {\kern 1pt} {\kern 1pt} {\kern 1pt} {\kern 1pt} {\kern 1pt} {\kern 1pt} {\kern 1pt} {\kern 1pt} {\kern 1pt} {\kern 1pt} {\kern 1pt} {\kern 1pt} {\kern 1pt} {\kern 1pt} {\kern 1pt} {\kern 1pt} {\kern 1pt} {\kern 1pt} {\kern 1pt} {\kern 1pt} {\kern 1pt} {\kern 1pt} {\kern 1pt} {\kern 1pt} {\kern 1pt}  < \underbrace {2\left( {\sum\limits_{m = 1}^{{N_R}} {{w_m}\left[ {(1 - \rho ){\gamma _m} - \rho } \right]} } \right)}_{Term{\kern 1pt} {\kern 1pt} {\kern 1pt} A}\underbrace {\left\{ \begin{array}{l}
\left[ {(1 - \rho ){\gamma _k} - \rho } \right] \cdot \sum\limits_{m \ne k} {w_m^2\left[ {{{(1 - \rho )}^2}{\gamma _m} + \frac{{1 + {\rho ^2}}}{2}} \right]} \\
{\kern 1pt} {\kern 1pt} {\kern 1pt} {\kern 1pt}  - \left[ {{{(1 - \rho )}^2}{\gamma _k} + \frac{{1 + {\rho ^2}}}{2}} \right] \cdot \sum\limits_{m \ne k} {w_m^2\frac{{w_k^o}}{{w_m^o}}\left[ {(1 - \rho ){\gamma _m} - \rho } \right]}
\end{array} \right\}}_{Term{\kern 1pt} {\kern 1pt} {\kern 1pt} B}\\
{\kern 1pt} {\kern 1pt} {\kern 1pt} {\kern 1pt} {\kern 1pt} {\kern 1pt} {\kern 1pt} {\kern 1pt} {\kern 1pt} {\kern 1pt} {\kern 1pt} {\kern 1pt} {\kern 1pt} {\kern 1pt} {\kern 1pt} {\kern 1pt} {\kern 1pt} {\kern 1pt} {\kern 1pt} {\kern 1pt} {\kern 1pt} {\kern 1pt} {\kern 1pt} {\kern 1pt} {\kern 1pt} {\kern 1pt} {\kern 1pt} {\kern 1pt}  = 0
\end{array}
\normalsize
\end{equation}
where we obtain the last equality by substituting $w_k^o$ and $w_m^o$ in accordance with (\ref{eq051}) into the second line.

Similarly, we can show that
\begin{equation}\label{eq055}
n({w_i}) > 0
\end{equation}

Thus, for an infinitesimally small $\varepsilon > 0$, we have that
\begin{equation}\label{eq056}
\frac{{\partial g({\bf{w}})}}{{\partial {w_i}}}\varepsilon  - \frac{{\partial g({\bf{w}})}}{{\partial {w_k}}}\varepsilon  = \varepsilon \left[ {\frac{{\partial g({\bf{w}})}}{{\partial {w_i}}} - \frac{{\partial g({\bf{w}})}}{{\partial {w_k}}}} \right] = \varepsilon \left[ {\frac{{n({w_i})}}{{d({w_i})}} - \frac{{n({w_k})}}{{d({w_k})}}} \right] > 0
\end{equation}

Given that $g({\bf{w}})$ is twice differentiable in ${w_k}$ and ${w_i}$, we can construct a feasible solution ${\bf{w'}}$ such that $g({\bf{w'}}) > g({\bf{w}})$ as follows:
\begin{equation}\label{eq057}
\left\{ \begin{array}{l}
{{w'}_k} = {w_k} - {\varepsilon  \mathord{\left/
 {\vphantom {\varepsilon  {\left| {\bf{K}} \right|}}} \right.
 \kern-\nulldelimiterspace} {\left| {\bf{K}} \right|}},{\kern 1pt} {\kern 1pt} {\kern 1pt} {\kern 1pt} {\kern 1pt} {\kern 1pt} {\kern 1pt} \forall k \in {\bf{K}}\\
{{w'}_i} = {w_i} + {\varepsilon  \mathord{\left/
 {\vphantom {\varepsilon  {\left| {\bf{I}} \right|}}} \right.
 \kern-\nulldelimiterspace} {\left| {\bf{I}} \right|}},{\kern 1pt} {\kern 1pt} {\kern 1pt} {\kern 1pt} {\kern 1pt} {\kern 1pt} {\kern 1pt} {\kern 1pt} {\kern 1pt} {\kern 1pt} {\kern 1pt} {\kern 1pt} {\kern 1pt} {\kern 1pt}{\kern 1pt} {\kern 1pt} \forall {\kern 1pt}  i  {\kern 1pt}  \in {\bf{I}}\\
{{w'}_j} = {w_j},{\kern 1pt} {\kern 1pt} {\kern 1pt} {\kern 1pt} {\kern 1pt} {\kern 1pt}  {\kern 1pt}  {\kern 1pt}  {\kern 1pt}  {\kern 1pt}  {\kern 1pt}  {\kern 1pt}  {\kern 1pt}  {\kern 1pt}  {\kern 1pt}  {\kern 1pt}  {\kern 1pt}  {\kern 1pt}  {\kern 1pt}  {\kern 1pt}  {\kern 1pt}  {\kern 1pt}  {\kern 1pt}  {\kern 1pt}  {\kern 1pt}  {\kern 1pt}  {\kern 1pt}  {\kern 1pt}  {\kern 1pt}  {\kern 1pt}  {\kern 1pt}  {\kern 1pt}  {\kern 1pt}  {\kern 1pt}  {\kern 1pt}  {\kern 1pt}  {\kern 1pt}  {\kern 1pt}  {\kern 1pt}  {\kern 1pt}  {\kern 1pt}  {\kern 1pt}  {\kern 1pt}  {\kern 1pt}  {\kern 1pt}  {\kern 1pt}  {\kern 1pt}  {\kern 1pt}  {\kern 1pt}  {\kern 1pt} {\kern 1pt} \forall  {\kern 1pt} j \in {\bf{J}}
\end{array} \right.
\end{equation}
where ${\bf{J}}=\left\{ {1,...,{N_R}} \right\} - {\bf{I}} - {\bf{K}}$ is the complement set of ${\bf{I}} \cup {\bf{K}}$; $\left| {\bf{I}} \right|$ and $\left| {\bf{K}} \right|$ denote the cardinality of ${\bf{I}}$ and ${\bf{K}}$, respectively.
\end{NewProof}

\begin{rem}\label{remark:1}
Although the above proof is complete by itself, a question that we might ask is how large can $\varepsilon$ be (i.e., it does not have to be infinitesimally small). By similar reasoning as in the proof, we note that as we increase $\varepsilon $, we would still have ${{\partial g({\bf{w'}})} \mathord{\left/ {\vphantom {{\partial g({\bf{w'}})} {\partial {{w'}_k}}}} \right. \kern-\nulldelimiterspace} {\partial {{w'}_k}}} < 0$ and ${{\partial g({\bf{w'}})} \mathord{\left/ {\vphantom {{\partial g({\bf{w'}})} {\partial {{w'}_i}}}} \right. \kern-\nulldelimiterspace} {\partial {{w'}_i}}} > 0$ provided that ${w_k}^\prime  > \left( {{{{w_j}} \mathord{\left/ {\vphantom {{{w_j}} {w_j^o}}} \right. \kern-\nulldelimiterspace} {w_j^o}}} \right)w_k^o$ for all $j \notin {\bf{K}}$ and ${w_i}^\prime  < \left( {{{{w_j}} \mathord{\left/ {\vphantom {{{w_j}} {w_j^o}}} \right. \kern-\nulldelimiterspace} {w_j^o}}} \right)w_i^o$ for all $j \notin {\bf{I}}$.

Thus, we can increase $\varepsilon$ until either ${w_k}^\prime  = \left( {{{{w_j}} \mathord{\left/  {\vphantom {{{w_j}} {w_j^o}}} \right. \kern-\nulldelimiterspace} {w_j^o}}} \right)w_k^o$ for some $j \notin {\bf{K}}$ (i.e., ${{{w_j}} \mathord{\left/ {\vphantom {{{w_j}} {w_j^o}}} \right. \kern-\nulldelimiterspace} {w_j^o}}$ is the second largest weight ratio here), or ${w_i}^\prime  = \left( {{{{w_j}} \mathord{\left/ {\vphantom {{{w_j}} {w_j^o}}} \right. \kern-\nulldelimiterspace} {w_j^o}}} \right)w_i^o$ for some $j \notin {\bf{I}}$ (i.e., ${{{w_j}} \mathord{\left/ {\vphantom {{{w_j}} {w_j^o}}} \right. \kern-\nulldelimiterspace} {w_j^o}}$ is the second smallest weight ratio here), whichever equality is fulfilled first. In particular, we can set

\begin{equation}\label{eq058}
\varepsilon  = \left\{ {\begin{array}{*{20}{c}}
{\min \left\{ {\left| {\bf{K}} \right|\left[ {{w_k} - \mathop {\max }\limits_{j \notin {\bf{K}}} \left( {\frac{{{w_j}}}{{w_j^o}}w_k^o} \right)} \right],{\kern 1pt} {\kern 1pt} {\kern 1pt} {\kern 1pt} {\kern 1pt} {\kern 1pt} {\kern 1pt} {\kern 1pt} \left| {\bf{I}} \right|\left[ {\mathop {\min }\limits_{j \notin {\bf{I}}} \left( {\frac{{{w_j}}}{{w_j^o}}w_i^o} \right) - {\kern 1pt} {w_i}} \right]{\kern 1pt} } \right\}}&{\rm{if} {\kern 1pt} {\kern 1pt}{\kern 1pt} {\kern 1pt} {\bf{J}} \ne \varnothing }\\
{\left| {\bf{K}} \right|({w_k} - {\kern 1pt} w_k^o) = \left| {\bf{I}} \right|({\kern 1pt} w_i^o - {w_i}){\kern 1pt} }&{\rm{if} {\kern 1pt} {\kern 1pt}{\kern 1pt} {\kern 1pt}{\bf{J}} = \varnothing }
\end{array}} \right.
\end{equation}

In fact, the above suggests an algorithmic way to march toward ${{\bf{w}}^o}$ from an arbitrary feasible ${\bf{w}}$. We perform (\ref{eq057}) in accordance with (\ref{eq058}). Then, with the new ${\bf{w'}}$, the cardinality of the new ${\bf{I}}$ or the new ${\bf{K}}$ is enlarged. We repeat the procedure until we get an even better solution ${\bf{w''}}$, The procedure is repeated until we reach ${{\bf{w}}^o}$.

\end{rem}

\section{Multi-antenna Packet Detection: Discussion and Experiments in a Distributed Antenna System}\label{sec-V}
Section IV puts forth two weight-assignment solutions for the combination of $r[n]$: (i) WFA and (ii) WMD. In a random-access network with co-located antennas, WFA and WMD have similar performance because co-located antennas have nearly the same SNR, resulting in similar weights for both WFA and WMD.\footnote{Signal at different co-located antennas may differ in phase, but it does not affect the calculation of ${a_R}[n]$ and $b[n]$. There is little SNR difference between co-located antennas. Hence, the weight assigned by WMD should be very close to that of WFA.} That is, WMD also results in roughly equal-weight assignments.

Packet detection in advanced wireless communication systems with distributed antennas, also known as distributed antenna systems (DAS), introduces different scenarios when comparing WFA and WMD because the SNRs at non-co-located antennas may vary widely. DAS offers two distinct advantages over conventional co-located antenna systems. First, co-located antennas suffer from a weakness that the signal blockage between the transmitter antenna and the co-located receiver antennas results in no signal reception. In contrast, DAS allows for potential signal reception even if one receiver antenna is blocked, thanks to clear paths of other non-blocked antennas. Second, in DAS, the proximity between the transmitter and the nearest receiver antenna tends to be smaller than the distance between the transmitter and co-located receiver antennas, resulting in improved communication quality between the transmitter and the receiver. However, the distributed nature of DAS poses a new challenge in weight assignment. Antennas in DAS can be separated by tens to hundreds of wavelengths. This discrepancy in propagation length among transmit-receive antenna pairs leads to varying SNRs across the antennas. Given the varying SNRs, the benchmarking of WFA and WMD becomes an issue.

\subsection{Implementation Issues}
Implementation-wise, WFA is a simple and practical scheme, as it requires no additional system information except $N_R$, the number of antennas. As a result, WFA has lower implementation complexity, requires fewer computational resources, and consumes less signal processing time. In particular, WFA does not need knowledge of the  antenna SNRs of the antennas since the weights do not depend on the SNRs.

For the false-alarm probability of WFA, we substitute ${w_j} = {1 \mathord{\left/ {\vphantom {1 {{N_R}}}} \right. \kern-\nulldelimiterspace} {{N_R}}}$ into (\ref{eq032}) and (\ref{eq033}) and obtain
\begin{equation}\label{eq059}
E\left( {r_N^{WFA}[n]} \right) =  - \rho {\kern 1pt} {\kern 1pt} {\kern 1pt} {\kern 1pt} {\kern 1pt} {\kern 1pt} {\kern 1pt} {\kern 1pt} {\kern 1pt} {\kern 1pt} {\kern 1pt} {\kern 1pt} {\kern 1pt} {\kern 1pt} {\kern 1pt} {\kern 1pt} {\kern 1pt} {\kern 1pt} {\kern 1pt} {\kern 1pt} {\kern 1pt} {\kern 1pt} {\kern 1pt} {\kern 1pt} {\rm{and}}{\kern 1pt} {\kern 1pt} {\kern 1pt} {\kern 1pt} {\kern 1pt} {\kern 1pt} {\kern 1pt} {\kern 1pt} {\kern 1pt} {\kern 1pt} {\kern 1pt} {\kern 1pt} {\kern 1pt} {\kern 1pt} {\kern 1pt} {\kern 1pt} {\kern 1pt} {\kern 1pt} {\kern 1pt} {\kern 1pt} {\kern 1pt} {\kern 1pt} {\kern 1pt} {\kern 1pt} Var\left( {r_N^{WFA}[n]} \right) = \frac{{1 + {\rho ^2}}}{{2\eta {N_R}}}
\end{equation}
where the subscript $N$ denotes pure noise input. We know from (\ref{eq059}) that the false-alarm probability of WFA can be written as
\begin{equation}\label{eq060}
P_{FA}^{WFA} = Q\left( { - \frac{{E\left( {r_N^{WFA}[n]} \right)}}{{\sqrt {Var\left( {r_N^{WFA}[n]} \right)} }}} \right) = Q\left( {\sqrt {\frac{{2\eta {N_R}{\rho ^2}}}{{1 + {\rho ^2}}}} } \right)
\end{equation}

For the missed-detection probability, we substitute ${w_j} = {1 \mathord{\left/ {\vphantom {1 {{N_R}}}} \right. \kern-\nulldelimiterspace} {{N_R}}}$ into (\ref{eq037}) and (\ref{eq038}) and obtain
\begin{equation}\label{eq061}
\left\{ \begin{array}{l}
E\left( {r_P^{WFA}[n]} \right) = \sum\limits_{j = 1}^{{N_R}} {{w_j}\left[ {(1 - \rho ){\gamma _j} - \rho } \right]}  = \left( {\frac{{1 - \rho }}{{{N_R}}}\sum\limits_{j = 1}^{{N_R}} {{\gamma _j}} } \right) - \rho \\
Var\left( {r_P^{WFA}[n]} \right) = \sum\limits_{j = 1}^{{N_R}} {w_j^2 \cdot \left[ {\frac{{{{(1 - \rho )}^2}}}{\eta }{\gamma _j} + \frac{{1 + {\rho ^2}}}{{2\eta }}} \right]}  = \left( {\frac{{{{(1 - \rho )}^2}}}{{\eta {N_R}}}\sum\limits_{j = 1}^{{N_R}} {{\gamma _j}} } \right) + \frac{{1 + {\rho ^2}}}{{2\eta {N_R}}}
\end{array} \right.
\end{equation}
where the subscript $P$ denotes packet-plus-noise input. From (\ref{eq061}), we have
\begin{equation}\label{eq062}
\small
\begin{array}{l}
P_{MD}^{WFA} = Q\left( {\frac{{E\left( {r_P^{WFA}[n]} \right)}}{{\sqrt {Var\left( {r_P^{WFA}[n]} \right)} }}} \right) = Q\left( {\sqrt \eta   \cdot {{\left( {\frac{{1 - \rho }}{{{N_R}}}\sum\limits_{j = 1}^{{N_R}} {{\gamma _j}}  - \rho } \right)} \mathord{\left/
 {\vphantom {{\left( {\frac{{1 - \rho }}{{{N_R}}}\sum\limits_{j = 1}^{{N_R}} {{\gamma _j}}  - \rho } \right)} {\sqrt {\frac{{{{(1 - \rho )}^2}}}{{N_R^2}}\sum\limits_{j = 1}^{{N_R}} {{\gamma _j}}  + \frac{{1 + {\rho ^2}}}{{2{N_R}}}} }}} \right.
 \kern-\nulldelimiterspace} {\sqrt {\frac{{{{(1 - \rho )}^2}}}{{N_R^2}}\sum\limits_{j = 1}^{{N_R}} {{\gamma _j}}  + \frac{{1 + {\rho ^2}}}{{2{N_R}}}} }}} \right)
\end{array} 
\normalsize
\end{equation}

Implementing WMD is more complex, as WMD requires knowledge of SNRs beforehand in order to calculate the weight of different antennas. However, obtaining the SNRs before packet detection is challenging, as accurate SNR estimation typically requires pilot-based signal processing, which is triggered by packet detection rather than preceding it. Thus, we have a ``chicken-and-egg dilemma" where we need to know the precise SNRs before doing WMD, but typically SNR estimations happen after WMD.

A possible practical way to overcome the problem is to estimate the SNR using the preamble. Specifically, we can measure the power of the background noise when the receiver is idle. And then, as in (\ref{eq004}), we calculate $b[n]$ for each antenna. We can coarsely estimate the SNR of an antenna with the noise power and the $b[n]$ of that antenna. However, this approach has limitations. First, the power of background noise may vary over time, but we only have the average noise power obtained during the idle period. Second, the weight assignment in this scheme is highly sensitive to interference. In shared spectrum environments, where most random-access systems are deployed, wireless interferences are common (say wireless packets from a Bluetooth device or a working microwave oven). These interferences may add to the preamble sequence and increase $b[n]$ (but such interference does not increase $a_R[n]$ or aid in the packet detection process), making the estimated SNR larger than its actual value. Consequently, this scheme may not be very reliable in practice.

Let us refer to the above system with the coarse SNR estimation as the practical WMD (P-WMD), and the hypothetical system with perfect \textit{a priori} knowledge of SNRs without estimation as the ideal WMD (I-WMD).

\subsection{Benchmarking WFA and WMD in two typical DAS scenarios}
This subsection benchmarks WFA and I-WMD/P-WMD with typical DAS scenarios (see Section \ref{sec-III} for the Pareto benchmarking method). In our benchmarking exercise, we set the maximum false alarm tolerance and missed detection tolerance at $P_{FA}^{\max } = {10^{ - 6}}$ and $P_{MD}^{\max } = {10^{ - 4}}$, respectively. That is, we are only interested in operating regions with false-alarm and missed-detection probabilities below these thresholds. We justify the tolerance settings in the following. In a practical DAS, it is reasonable to assume that a packet has no more than 1024 OFDM samples.\footnote{In recent WiFi standards such as 802.11ax or future IEEE 802.11be, the length of a packet is typically no larger than 1024 samples. Additionally, a recent technical trend in wireless communication is to achieve URLLC with very short packet lengths. This setting is referred to as short-packet communication (SPC), where a packet typically has no more than 50 bytes \cite{ref22}, making the packet length much shorter than 1024 samples.} Suppose that we want the system to experience no more than one false alarm every 1000 packets on average. Then, the false-alarm probability should be no larger than ${\left( {1024 \times 1000} \right)^{ - 1}} \approx {10^{ - 6}}$. As for the missed-detection probability, we take as reference the ultra-reliable low latency communication (URLLC) defined by 3rd generation partnership project (3GPP) that requires at least 99.99\% successful packet decoding \cite{ref23,ref24}, i.e., the transmission error tolerance is no larger than 0.01\% or $10^{-4}$. We assume a missed-detection tolerance commensurate with the reliability requirement of packet decoding, corresponding to a missing-detection probability of no more than ${10^{-4}}$.

Furthermore, as in the previous analysis, we consider a two-STS preamble, with each STS having 16 samples (i.e., $\eta = 16$). We further assume that the timing offsets between different packets have been compensated prior to the application of WFA/I-WMD/P-WMD.\footnote{Sample misalignment can be a challenge for DAS. Due to the path-length discrepancy between different transmit-receive antenna pairs, the samples collected at different antennas may not be aligned in time (i.e., different propagation latency at different antennas). Therefore, the alignment of input samples is necessary before applying WFA/WMD in DAS. For preambles with two STSs, the $b[n]$ of each packet exhibits a sharp peak that indicates the clear starting position of the preamble (see Fig. \ref{fig:2}). We can utilize the peak for alignment.}

We consider two typical DAS scenarios in the benchmark: the non-blocked scenario and the partially blocked scenario. In the non-blocked scenario, all transmit-receive pairs have clear propagation paths. In the partially blocked scenario, the propagation paths of some receive antennas are obstructed, leading to significantly lower SNRs than other antennas. We do not consider the case where all receiver antennas are blocked because it is unlikely to happen (since DAS is designed to avoid such situations). Furthermore, even if the rare occasion happens, improving packet-detection performance would be futile as the low SNRs in all antennas could prevent successful packet decoding anyway.

Table I gives two examples of the SNR conditions in the non-blocked scenario and the partially blocked scenario. Both examples consider four distributed antennas.
\begin{table}[htbp]\label{table:1}
\captionsetup{font={small}}
\caption{SNR conditions of two typical DAS examples.}
\vspace{-2em}
\begin{center}
\begin{tabular}{|p{1.8cm}<{\centering}|p{1.4cm}<{\centering}|p{1.4cm}<{\centering}|p{1.4cm}<{\centering}|p{1.4cm}<{\centering}|p{1.4cm}<{\centering}|p{1.4cm}<{\centering}|p{1.4cm}<{\centering}|p{1.4cm}<{\centering}|}
\hline
{\multirow{2}{*}{}} & \multicolumn{4}{c|}{Example One (Non-blocked DAS)} & \multicolumn{4}{c|}{Example Two (Partially Blocked DAS))}\\
\cline{2-9}
&Antenna 1&Antenna 2&Antenna 3&Antenna 4&Antenna 1&Antenna 2&Antenna 3&Antenna 4\\
\hline
SNR/dB&3.6118&3.8903&4.0338&3.3649&0.2013&3.5843&3.3318&4.2489\\
\hline
\end{tabular}
\end{center}
\end{table}

\begin{figure}[htbp]
  \centering
  \subfloat[]  {\label{fig:11} \includegraphics[width=0.45\textwidth]{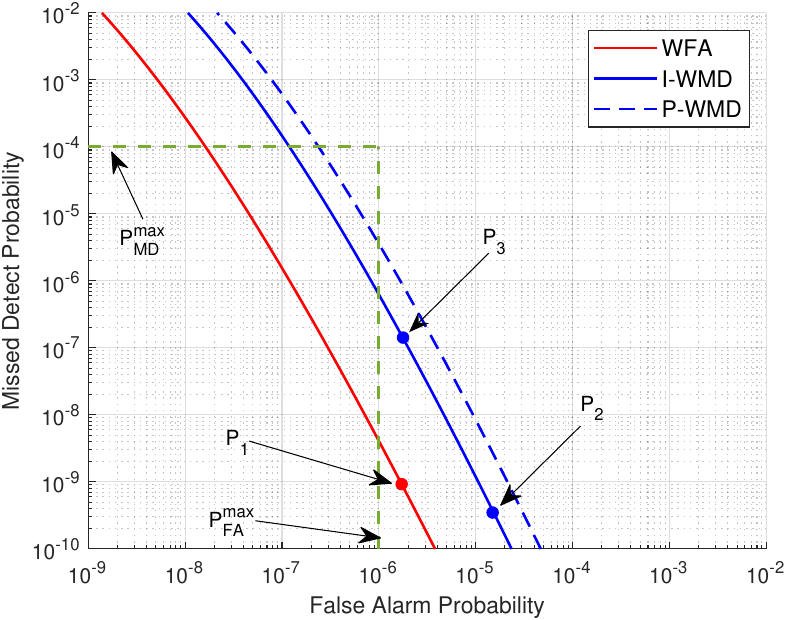}}
  \quad \quad
  \subfloat[]  {\label{fig:12} \includegraphics[width=0.45\textwidth]{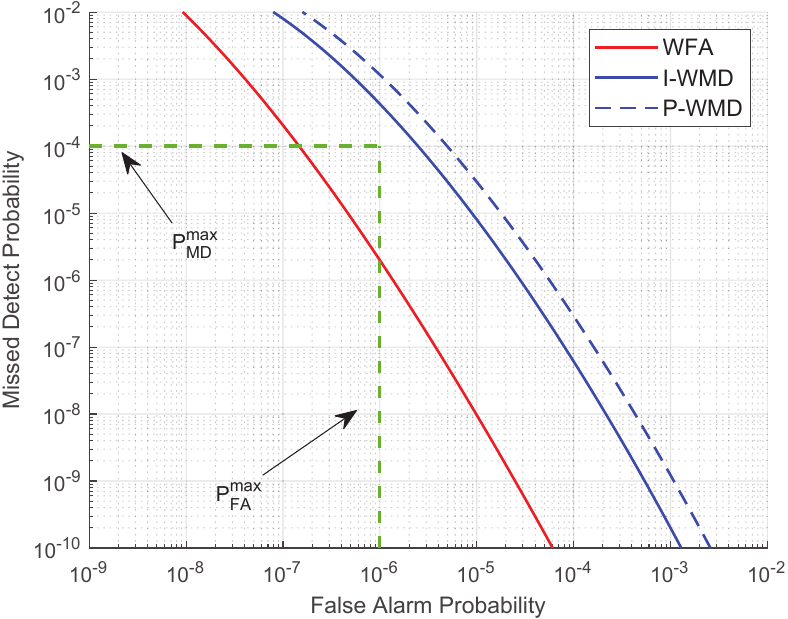}}
  \captionsetup{font={small}}
  \caption{(a) MD-FA curves of WFA/I-WMD/P-WMD in the non-blocked scenario. (b) MD-FA curves of WFA/I-WMD/P-WMD in the partially blocked scenario.}
\end{figure}

Fig. \ref{fig:11} presents the MD-FA curves of WFA, I-WMD, and P-WMD in the non-blocked scenario. It is clear that WFA outperforms the other two schemes in this example. The reader may wonder why the two WMD schemes turn out to have inferior missed-detection performance than WFA. The reason is simple: WMD is superior to WFA in terms of missed-detection performance only for a given fixed detection threshold $\rho$. However, WMD has a higher false-alarm probability for that fixed $\rho$. For example, in Fig. \ref{fig:11}, we fix $\rho=0.45$ and highlight the corresponding $\left(P_{FA},P_{MD}\right)$ for WFA and I-WMD in points $P_1$ and $P_2$, respectively. As the figure shows, although $P_2$ has a lower missed-detection probability, its false-alarm probability is much higher than $P_1$. For the same false-alarm performance as WFA, I-WMD would have to raise its $\rho$, which in turn increases its missed-detection probability, to the extent that it is now worse than that of WFA (see $P_3$ in the figure).

For the partially blocked scenario, as we can see from Table I, antenna one experiences a significantly low SNR due to blockage, while the other three antennas remain unaffected. Fig. \ref{fig:12} presents MD-FA curves of WFA, I-WMD, and P-WMD for this example. It is clear that WFA still outperforms the other two schemes in the partially blocked scenario.

\subsection{ General Benchmark Results in a Distributed Antenna Dataset}
After examining the above two typical scenarios, we now proceed to a more general comparison between WFA and I-WMD/P-WMD. We use the same benchmark scheme as in subsection B (including the same $P_{FA}^{\max }$ and $P_{MD}^{\max }$ settings) and conduct emulation experiments on DICHASUS \cite{dichasus}, a massive open-source wireless channel dataset collected in industrial environments. To conserve space, we do not present the numerous MD-FA curves here.

\begin{figure}[htbp]
  \centering
  \includegraphics[width=0.45\textwidth]{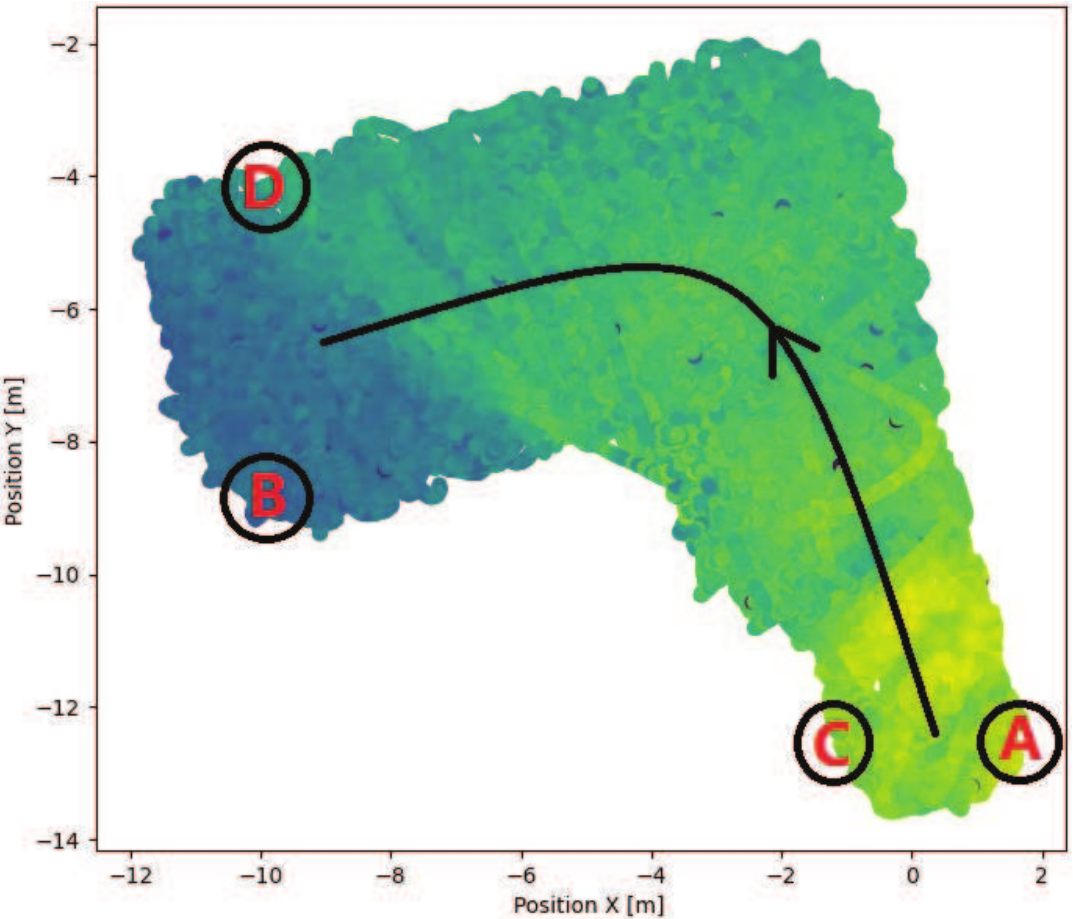}\\
  \captionsetup{font={small}}
  \caption{The location of Group A/B/C/D and an example of a moving transmitter.}
\label{fig:15}
\end{figure}

We first give a general introduction of the dataset. The channel information in DICHASUS was measured using 32 software-defined radio (SDR) sensors and one transmitter that moves randomly in a factory. These 32 sensors were divided into four groups (Group A, B, C, and D), with each group comprising eight sensors located in one corner of the factory. Fig. \ref{fig:15} shows the layout of the factory and the locations of the antenna groups A, B, C, and D. The transmitter periodically transmits a reference packet that is known to every sensor. Upon receiving the reference packet, an SDR sensor compares it with the original version of the packet it knows \textit{a priori} to obtain the precise channel information. The DICHASUS dataset encompasses in total 44,703 valid\footnote{In the data pre-processing stage, we discard a small portion of measurements that are obviously invalid (or even wrong). For example, SNRs of some antennas are unreadable (i.e., not a number, NaN) or smaller than 0dB. That may be caused by measurement (or data recording) errors during the data collection. Furthermore, studying cases with less than 0dB SNR is meaningless for our packet-detection research, as such cases will fail in packet decoding anyway.} DAS measurements collected in five different days, and each measurement has 32 pieces of channel information estimated by the 4x8 distributed sensors through the same reference packet at the same time.

We next give a detailed look at the SNR information in DICHASUS. We denote an antenna by $Ana(i,j)$, where $i \in \left\{ {A,B,C,D} \right\}$ and $j \in \left\{ {1,2,3,4,5,6,7,8} \right\}$. Each $Ana(i,j)$ has 44,703 SNR measurements, and we denote the ${k^{th}}$ SNR measurement by ${\gamma _{i,j}}(k)$. Unless stated otherwise, experiments and discussions below assume the original SNR values rather than the dB values.

With the above definition, we study the SNR correlation of all 32 antennas and present the result in Table II below. Without loss of generality, we use $Ana(A,1)$ as a reference and calculate the correlation between $Ana(A,1)$ and $Ana(i,j)$ by
\begin{equation}\label{eq063}
corr\left\{ {(A,1),(i,j)} \right\} = \frac{{\sum\nolimits_k {\left( {{\gamma _{A,1}}(k) - \overline {{\gamma _{A,1}}} } \right)\left( {{\gamma _{i,j}}(k) - \overline {{\gamma _{i,j}}} } \right)} }}{{\sqrt {\sum\nolimits_k {{{\left( {{\gamma _{A,1}}(k) - \overline {{\gamma _{A,1}}} } \right)}^2} \cdot \sum\nolimits_k {{{\left( {{\gamma _{i,j}}(k) - \overline {{\gamma _{i,j}}} } \right)}^2}} } } }}
\end{equation}

\begin{table}[htbp]\label{table:2}
\captionsetup{font={small}}
\caption{SNR correlations among different antennas (using $Ana(A,1)$ as the reference).}
\vspace{-2em}
\begin{center}
\begin{tabular}{|p{1.25cm}<{\centering} | p{1.0cm}<{\centering} | p{1.0cm}<{\centering} | p{1.0cm}<{\centering} | p{1.0cm}<{\centering} | p{1.0cm}<{\centering} | p{1.0cm}<{\centering} | p{1.0cm}<{\centering} | p{1.0cm}<{\centering}|}
\hline
&$j=1$&$j=2$&$j=3$&$j=4$&$j=5$&$j=6$&$j=7$&$j=8$\\
\hline
Group A&1.0000&0.8632&0.9403&0.8833&0.9232&0.8963&0.8611&0.8228\\
\hline
Group B&-0.2790&-0.3556&-0.2889&-0.2540&-0.2879&-0.3207&-0.3787&-0.3002\\
\hline
Group C&0.4093&0.4889&0.4658&0.5190&0.5212&0.4353&0.4197&0.4866\\
\hline
Group D&-0.4137&-0.3966&-0.3928&-0.3709&-0.4407&-0.4583&-0.4009&-0.4190\\
\hline
\end{tabular}
\end{center}
\end{table}

\begin{figure}[htbp]
  \centering
  \subfloat[]  {\label{fig:13} \includegraphics[width=0.45\textwidth]{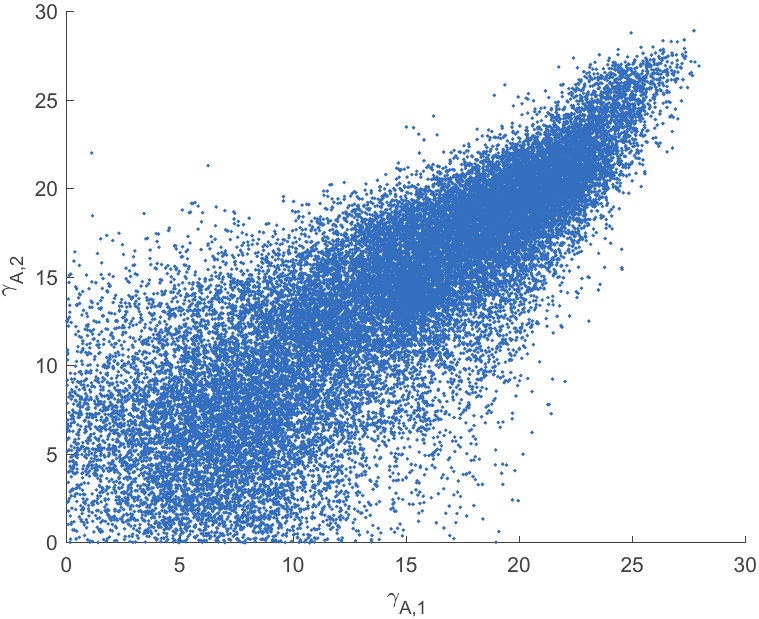}}
  \quad \quad
  \subfloat[]  {\label{fig:14} \includegraphics[width=0.45\textwidth]{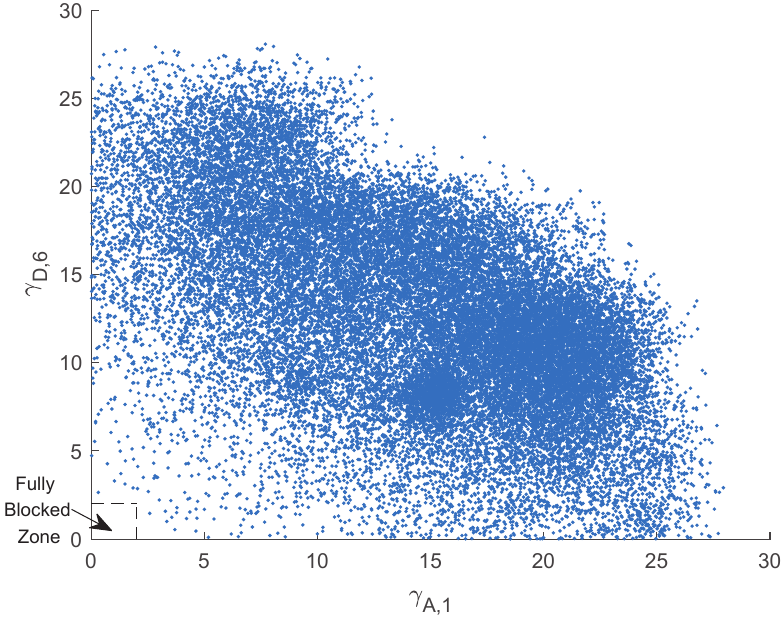}}
  \captionsetup{font={small}}
  \caption{(a) The scatter plot of SNRs in $Ana(A,1)$ and $Ana(A,2)$. (b) The scatter plot of SNRs in $Ana(A,1)$ and $Ana(D,6)$. In (b), the ``fully blocked zone"  refers to the SNR region where both two antennas have very low SNR. Both two figures use the original SNR values rather than the dB values.}
\end{figure}

We have several observations from Table II. First, SNRs of two co-located antennas are highly positively correlated. For example, Fig. \ref{fig:13} plots a 2-D scatter chart for $Ana(A,1)$ and $Ana(A,2)$, where a scatter point $\left( {{\gamma _{A,1}}(k),{\kern 1pt} {\kern 1pt} {\kern 1pt} {\kern 1pt} {\gamma _{A,2}}(k)} \right)$ is plotted for each $k$. We see that the points roughly fall around the straight line $y = x$. We can obtain many similar 2-D scatter charts if we consider co-located antennas within the same group.

Second, if we look at $Ana(A,1)$ and antennas in Group D, we observe a weak and negative correlation between $Ana(A,1)$ and $Ana(D,j)$. Fig. \ref{fig:14} uses $Ana(D,6)$ as an example to illustrate the relationship. As we can see from the figure, the SNRs of $Ana(A,1)$ and $Ana(D,6)$ are generally negatively correlated. This can be explained by the distributed nature of DAS: if one distributed antenna is weak, the other may still be strong. From Fig. \ref{fig:15}, we see that Group A and Group D are located in two opposite corners of the factory. When the transmitter moves from one corner to the opposite corner (Fig. \ref{fig:15} also provides an example of the trace of a transmitter moving from Group A to Group D), we should observe a decrease in SNR for one and an increase in SNR for the other. Furthermore, thanks to this distributed nature of DAS, we see no fully blocked cases in Fig. \ref{fig:14}, i.e., at least one antenna has an SNR larger than 3dB.


We now elaborate on our experimental settings. In the first step, we construct a fixed two-STS preamble (with $\eta=16$ in every STS). This preamble sequence is used consistently throughout the experiment. Since our objective is to benchmark WFA and I-WMD/P-WMD in general cases to see if one scheme consistently outperforms the other in practice, we emulate and test all channel measurements in DICHASUS. For the testing of each measurement, it is sufficient to represent a group of antennas with one or two elements, given that the SNRs of co-located antennas within the same group are highly positively correlated. Therefore, we randomly select two antennas from each group to simplify the emulation. After the random antenna selections, we emulate the corresponding 4x2 channels and transmit the fixed preamble sequence through these emulated channels. At the receiver side of the eight-antenna system, we apply WFA/I-WMD/P-WMD and record the benchmark result. After testing one measurement, we move on to the next measurement until we have completed testing all 44,703 measurements in the DICHASUS dataset.

During the benchmark process, we find that some antennas have significantly lower SNR than the other antennas when blockages occur. Let us make the following definition: if an antenna's SNR is lower than 3dB, we say the antenna is blocked and the measurement is a blocked case. Our analysis reveals that 14.47\% of the emulated eight-antenna systems encounter varying degrees of blockage, while the remaining 85.53\% of the emulated systems are non-blocked. Thanks to the reliability advantage of DAS, no fully blocked cases (all antennas blocked) is observed in the dataset.

Table III presents emulation results in detail, with 14.47\% partially blocked tested cases and 85.53\% non-blocked cases. In the table, each row corresponds to the benchmark conducted on a specific day. For each day, there are three pieces of data that record the percentage of instances where ``WFA outperforms I-WMD/P-WMD in terms of packet detection", ``Draw", and ``I-WMD/P-WMD outperforms WFA" (please refer to Section \ref{sec-III} for the explanation of ``Draw"). The last row of the table provides the average results across all 44,703 measurements.
\begin{table}[htbp]\label{table:3}
\captionsetup{font={small}}
\caption{Benchmarking WFA and I-WMD/P-WMD with the whole DICHASUS dataset.}
\vspace{-2em}
\begin{center}
\begin{tabular}{|p{1.0cm}<{\centering}|p{1.8cm}<{\centering}|p{1.8cm}<{\centering}|p{1.8cm}<{\centering}|p{1.8cm}<{\centering}|p{1.8cm}<{\centering}|p{1.8cm}<{\centering}|}
\hline
{\multirow{2}{*}{}} & \multicolumn{3}{c|}{Benchmark One (WFA v.s. I-WMD)} & \multicolumn{3}{c|}{Benchmark Two (WFA v.s. P-WMD)}\\
\cline{2-7}
&WFA wins & Draw & I-WMD wins & WFA wins & Draw & P-WMD wins\\
\hline
Day 1& 0.9218 & 0.0431 & 0.0351 & 0.9902 & 0.0048 & 0.0050\\
\hline
Day 2& 0.9002 & 0.0646 & 0.0352 & 0.9843 & 0.0079 & 0.0078\\
\hline
Day 3& 0.9104 & 0.0689 & 0.0207 & 0.9858 & 0.0101 & 0.0041\\
\hline
Day 4& 0.9169 & 0.0689 & 0.0142 & 0.9821 & 0.0107 & 0.0072\\
\hline
Day 5& 0.9251 & 0.0481 & 0.0268 & 0.9888 & 0.0058 & 0.0054\\
\hline
Avg. & 0.9149 & 0.0587 & 0.0264 & 0.9862 & 0.0079 & 0.0059\\
\hline
\end{tabular}
\end{center}
\end{table}

Table III shows that WFA outperforms both P-WMD and I-WMD by a significant margin. On average, WFA surpasses its opponents in 91.49\% of cases when competing against I-WMD and in 98.62\% of cases when competing against P-WMD. Further, considering the draw cases, the percentage of WFA not losing is 97.36\% when competing with I-WMD and 99.41\% when competing with P-WMD.

Based on the above emulation results, we recommend WFA as the desirable choice for a realistic DAS due to the following reasons:
\begin{enumerate}[1)]
    \item (Superior packet-detection performance) WFA consistently outperforms both I-WMD and P-WMD in most cases, with an average percentage of 97.36\% and 99.41\%, respectively.
    \item (Simplicity) WFA is much easier to implement, as it does not require SNR estimations or complex weight calculations. In contrast, I-WMD is not practical implementation-wise.
    \item (Reliability) WFA is more reliable than P-WMD because it is not sensitive to noise or interference, which is critical in practical applications where environmental factors can affect signal quality.
\end{enumerate}

\section{Conclusion}\label{sec-VI}
In conclusion, this paper has provided a comprehensive treatment of packet detection for random access networks. The conventional S\&C algorithm suffers from complex correlated noises in its packet-detection metric, making it difficult to analyze. To address this issue, we propose an analytical framework that uses ``compensated autocorrelation" as the new metric for packet detection. In addition, our results demonstrate that taking the real part of the autocorrelation can significantly enhance the performance of S\&C.

By leveraging the analytical tractability of compensated autocorrelation, we obtain accurate closed-form expressions for false-alarm and missed-detection probabilities. These expressions provide a rigorous theoretical foundation for fair Pareto benchmarking of packet-detection schemes and extension of single-antenna packet detection schemes to multi-antenna packet detection schemes.

In particular, for multi-antenna detection, we can use the weighted sum of compensated autocorrelations at different antennas as the metric without sacrificing analytical rigor. This approach enables us to determine the best weights for minimizing the false-alarm probability (WFA) and the missed detection probability (WMD). Our investigation suggests that WFA is the preferred choice for practical application settings.

Overall, our paper contributes to both the theory and practice of packet detection for random access networks. Our theoretical foundation provides insights on how to design packet detection schemes and how to compare and benchmark them in a rigorous manner in practical systems. This work has the potential to improve the performance of packet detection in random access networks and advance the field toward more efficient and reliable communication systems.

\section{Acknowledgment}
The authors would like to express their sincere gratitude to Prof. Henry Chen for his valuable suggestion to use the DICHASUS dataset for our emulation experiments. The authors would also like to extend their appreciation to Dr. Gongpu Chen for his helpful comments on the Q function derivation in this paper.

\bibliographystyle{IEEEtran}
\bibliography{./Reference/Ref_IEEE}

\end{document}